\documentclass[useAMS]{mn2e}
\usepackage{graphicx}
\usepackage{lscape}

\def\lesssim{\mathrel{\hbox{\rlap{\hbox{\lower4pt\hbox{$\sim$}}}\hbox{$<$}}}}

\def\gtrsim{\mathrel{\hbox{\rlap{\hbox{\lower4pt\hbox{$\sim$}}}\hbox{$>$}}}}

\title[The age-metallicity dependence for white dwarfs]
{The age-metallicity dependence for white dwarfs stars}

\author[A. D. Romero, F. Campos and S. O. Kepler]
{A. D. Romero$^{1}\thanks{E-mail: alejandra.romero@ufrgs.br}$, F. Campos$^{1}$ and S. O. Kepler$^{1}$
\newauthor\\
 $^{1}$Departamento de Astronomia, Universidade Federal do 
Rio Grande do Sul, Av. Bento Goncalves 9500
Porto Alegre 91501-970, RS, Brazil\\}

\begin{document}
\date{}

\maketitle

\label{firstpage}

\begin{abstract}
We present  a theoretical study  on the metallicity dependence  of the
initial$-$to$-$final    mass   relation    and   its    influence    on  white dwarf  age
determinations. 
We  compute a grid of evolutionary  sequences from the
main sequence  to $\sim 3\, 000$ K on the  white dwarf cooling  curve, passing
through all intermediate stages. During the thermally-pulsing asymptotic giant branch no third dredge-up episodes are considered 
and thus the photospheric C/O ratio is below unity for sequences with metallicities larger than $Z=0.0001$. We  consider initial
metallicities from  $Z=0.0001$  to  $Z=0.04$, accounting  for  stellar
populations in  the galactic disk  and halo, with initial  masses below
$\sim 3M_{\odot}$.   We found a clear  dependence of the  shape of the
initial$-$to$-$final  mass relation  with the  progenitor  metallicity, where
 metal  rich  progenitors  result  in less  massive  white  dwarf
remnants, due to  an enhancement of the mass  loss rates associated to
high  metallicity values.  By  comparing our  theoretical computations
with semi empirical data from globular and old open clusters, we found
that the observed intrinsic mass spread can be accounted for by a set
of initial$-$to$-$final mass relations characterized by different metallicity
values.  Also, we  confirm that the  lifetime spent before  the white dwarf stage  increases with
metallicity.  Finally, we estimate  the mean  mass at  the top  of the
white dwarf cooling curve for three globular clusters NGC 6397, M4 and
47 Tuc, around $0.53 M_{\odot}$, characteristic  of old stellar
populations.  However, we  found different  values for  the progenitor
mass, lower for  the metal poor cluster, NGC 6397,  and larger for the
younger  and  metal  rich  cluster   47  Tuc,  as  expected  from  the
metallicity dependence of the initial$-$to$-$final mass relation.
\end{abstract}



\section{Introduction}

White   dwarf  stars   are  the   most  common   stellar  evolutionary
endpoint. All  stars with initial  masses up to $\sim  10.5 M_{\odot}$
(Smartt 2009, Doherty et al. 2014), more than 97\% of the stars in the
Galaxy, end their lives as  white dwarf stars.  Their evolution can be
described as a  simple and slow cooling process.   Because white dwarf
star  are  abundant  and  long-lived  objects  they  convey  important
information about the properties  of all galactic populations (Isern  et al.  2001; 
Liebertet al.  2005; Bono et  al. 2013).  In particular,
white dwarf stars can be considered as reliable cosmic clocks to infer
the age of a wide variety of stellar populations, such as the Galactic
disks and halo  (Winget et al. 1987; Garc\'{i}a-Berro  et al.  1988ab,
Isern et al 1998; Torres et al.  2002), and globular and open clusters
(Kalirai et al. 2001; Hansen et al. 2002, 2007; Kalirai 2013).

Methods for  determining stellar population ages from  their white dwarf
cooling sequences  are usually based  on the comparison of  either the
observed white dwarf luminosity function (Winget et al. 1987, Bedin et
al. 2010,  Garc\'{i}a-Berro et al.  2014), or the distribution  in the
Color--Magnitude  Diagram,  with  theoretical computations  (see  e.g.
Hansen et  al.  2007).  Both techniques  rely on the  use of extensive
grids  of white  dwarf cooling  sequences.  However,  the evolutionary
stages prior to the final cooling curve must be taken into account.
In particular, a connection between  the properties of the white dwarfs
with  those of their  main sequence  progenitors is  the initial-to-final
mass  relation (IFMR).  This  relation  can be  applied  to study  the
chemical  evolution  of  galaxies,  including the  enrichment  of  the
interstellar  medium, and  the chemical  evolution history  of stellar
populations in  general.  At the high  mass end, the white dwarf IFMR  can lead to
constraints on the critical mass that separates white dwarf production
from  Type II  supernova  explosions, and  at  the low  mass end,  the
relation represents a tool to  probe the progenitor properties of most
of the evolved stars in old stellar populations, the majority of which
are now low-mass white dwarf stars.

The study  of the IFMR started  with the work of  Weidemann (1977) who
compared theoretical models  of mass loss to the  observed masses of a
few  white dwarfs  in  nearby stellar  clusters.  This  semi-empirical
approach was  followed by other  authors. More recently, Catal\'an  et al.
(2008a) obtained a semi-empirical  IFMR by re-evaluating the available
data,  that  includes observations  from  individual  stars from  open
clusters  and common  proper  motion pairs.   Kalirai  et al.   (2008)
increased the  observations and included  data from two  open clusters
older  than $\sim$  1 Gyr  and  from the  old open  cluster NGC  6791,
extending the  initial mass  range down to  1.16$M_{\odot}$.  Finally,
Kalirai (2013)  presented an IFMR based  on spectroscopic observations
of white dwarfs  belonging to open and globular  clusters and Sirius B
(see e.g. Kalirai et al.  2007, 2008, 2009), with initial mass ranging
from  $0.8M_{\odot}$  to  $\sim  6M_{\odot}$.   In  particular,  Kalirai et al. (2008)
 bin the relation so  that each star cluster was represented by
a single point.

In  the last few  years, the  amount of  data available  has increased
considerably.   In the Galactic  field, the  Sloan Digital  Sky Survey
increased the total number of spectroscopically confirmed white dwarfs
stars to $\sim 30\, 000$ (Kleinmann et al.  2013, Kepler et al. 2014).
In the  Galactic Halo, recent  Hubble Space Telescope  observations of
the globular  clusters M4 (Richer et  al.  2004; Hansen  et al.  2004;
Bedin et al. 2009), $\omega$ Cen (Monelli et al.  2005), NGC 6397 (Richer
et al.   2006; Hansen et  al. 2007) and  47 Tuc (Hansen et  al.  2013)
have similarly uncovered several hundreds of cluster white dwarfs.  In
all  cases,  the  white  dwarf  samples  are  dominated  by  low  mass
stars. With the  increase of the available data,  the general trend in
the  empirical IFMR  remains. However,  the  scatter in  the data  has
increased, possibly due  to a correlation between the  mass loss rates
and  the properties of  the host  environment.  Although  this scatter
could be related to other properties as rotation, binary evolution and
the presence  of magnetic fields  (Weidemann 2000), it is  more likely
produced by  differences in  the chemical composition,  i.e. different
metallicity values (Kalirai et al.  2008).

To  estimate  the age  of  a given  stellar population,  the
metallicity of the environment has to  be taken into account. As it is
well known, for a fixed initial mass, the lifetime of a star decreases
with  the  metallicity,  since  less  metallic  --  lower  opacity  --
envelopes leads to higher luminosities.  On the other  hand, an increase
on metallicity leads to an  enhanced mass loss during the giant phases
of the evolution, resulting in  lower final masses. In addition, other
parameter characterizing  white dwarf stars are expected  to depend on
metallicity.  For instance,  the amount  of hydrogen  left on  a white
dwarf star  decreases for higher metallicity values  (see e.g.  Renedo
et al.  2010). Also, a reduction  of the initial  metallicity tends to
reduce the  degree of chemical stratification in  the central regions,
which translates in a lower  oxygen abundance at the center (Dominguez
et al. 1999).  This will affect the cooling  rates, and in particular,
the amount  of energy released  by the crystallization process  at 
$ T_{\rm eff}\lesssim 12\, 000$ K.

In  this work  we  focus on  the  study of  the  effects of  different
metallicity values on the IFMR and on the lifetimes during the cooling
sequences  and  its  previous  evolutionary stages.   Specifically  we
center our  analysis on the low and intermediate mass end of the  IFMR, by considering
masses ranging  from $\sim 0.8M_{\odot}$  to $\sim 3M_{\odot}$  at the
main  sequence.  To this  end we  compute full  evolutionary sequences
from the  zero age  main sequence (ZAMS),  through the  central stable
hydrogen  and  helium  central  burning,  the  thermally  pulsing  and
mass-loss phases on the asymptotic  giant branch (AGB), and finally to
the white  dwarf cooling phase.  The computations  were performed with
the LPCODE evolutionary  code (Althaus et al.  2005a).   We consider 7
metallicity values  ranging from $Z=0.0001$ to  $Z=0.04$, covering the
values  for halo  and disk  populations.  

This  paper  is  organized  as follows.   Section  \ref{numerical}  is
devoted to  present the input  physics considered in  the evolutionary
computations.
In section \ref{parameters} we present  our model grid and discuss the
influence  of  metallicity  on  the  pre-white  dwarf  evolution. We 
explore the influence  of metallicity on the minimum mass
to  reach helium  burning at  the beginning  of the  horizontal branch
phase and  the core mass growth  during the TP-AGB phase.   In section
\ref{i-f} we present and discuss  our results for the theoretical IFMR
for  the  different  metallicity  values and  compare  them  with  the
observational data from  open and globular clusters.  Also, we compare
our results  with other theoretical and  semi-empirical determinations
of the IFMR.  In section \ref{wd-parameters} we  analyze the influence
of the  metallicity on  structural parameters  of the  resulting white
dwarf  models,  such  as  the  amount of  hydrogen  and  the  chemical
composition of  the core.  Section  \ref{age} is devoted to  asses the
effect   of  metallicity   on   the  age   determination  of   stellar
populations.  In  particular  we   use  our  theoretical  evolutionary
sequences to study three well known globular cluster, NGC 6397, M4 and
47 Tuc.   For this systems we  estimate the mass range  at the cooling
sequence.    We   present   our    concluding   remarks   in   section
\ref{conclusions}.

\section{Numerical tools}
\label{numerical}
 
The evolutionary  computations presented in this  work were calculated
with  an updated  version of  the  LPCODE evolutionary  code. The details on the code
can be  found in  Althaus et  al.  (2005a,  2010),  Renedo et
al. (2010) and Romero et al. (2013).  The LPCODE evolutionary code  
computes the complete evolution from the ZAMS,  through  the  hydrogen  
and helium  burning
stages, the thermally pulsating and mass loss stages on the asymptotic
giant branch, to the white dwarf cooling evolution.
This code has been used
to produce  very accurate white dwarf models  (see Garc\'{i}a-Berro et
al.  2010; Althaus  et  al. 2010;  Renedo  et al.  2010 and  reference
therein).  The code  has  also been  used  to study  the formation  of
extreme horizontal  branch stars (Miller  Bertolami et al.  2008), the
evolution  of extremely low  mass white  dwarf stars  (Althaus et al.
  2013), and also the role  of thermohaline mixing
for the surface composition of low-mass red giants (Wachlin et al. 2011).

\subsection{Pre-white dwarf evolution}
\label{pre-wd}

The  LPCODE considers simultaneous  treatment of
non-instantaneous  mixing  and burning  of  elements  (Althaus et  al.
2003).  The nuclear  network accounts for 16 elements  and 34 reaction
rates,  that  includes  $pp-$chain,  CNO$-$cycle, helium  burning  and
carbon ignition (Althaus et al.  2005b). Reaction rates are taken from
Caughlan \& Fowler  (1988) and Angulo et al.   (1999).  In particular,
the  $^{12}$C$(\alpha ,  \gamma )  ^{16}$O reaction  rate,  of special
relevance  for  the carbon$-$oxygen  stratification  of the  resulting
white  dwarf, was  taken from  Angulo et  al.  (1999).   The radiative
opacities are  those of  the OPAL project  (Iglesias \&  Rogers 1996),
including  carbon$-$ and  oxygen$-$rich compositions,  supplemented at
low  temperatures   with  the  molecular  opacities   of  Ferguson  et
al.  (2005)  and  Weiss  \&  Ferguson  (2009).   In  LPCODE  molecular
opacities are  computed adopting the  opacity tables with  the correct
abundances of  the unenhanced metals and  the appropriate carbon-oxygen
ratio. For the present computations, we have not considered carbon-enriched molecular opacities (Marigo 2002), 
which are expected to reduce effective temperatures at the AGB (Wiess \& Ferguson 2009). Conductive   opacities    are   those   from   Cassisi   et
al. (2007). Neutrino emission rates for pair, photo and bremsstrahlung
processes are taken from Itoh  et al. (1996), and for plasma processes
the treatment  of Haft et al.  (1994) is included. We adopted the standard
mixing  length theory  with the  free  parameter $\alpha  = 1.6$.  
With this value, the present luminosity
and effective temperature of the  Sun, $\log T_{\rm eff} = 3.7641$ and
$L_{\odot} =  3.842 \times  10^{33} {\rm erg\,  s^{-1}}$ at an  age of
4570 Myrs, are reproduced by  LPCODE when $Z=0.0164$ and $X=0.714$ are
adopted,  in agreement  with the  $Z/X$  value of  Grevesse \&  Sauval
(1998),  which are better reproduced by helioseismology than more recent 
determinations (Basu \& Antia 2008).

During  the evolutionary  stages prior  to the  thermally  pulsing AGB
phase, we  allow the occurrence of extra$-$mixing  episodes beyond each
convective  boundary  following  the  prescription of  Herwig  et  al.
(1997).  Note  that the  occurrence of extra$-$mixing  episodes during
the core  helium burning phase  largely determines the  final chemical
composition  of the  core of  the  resulting white  dwarf star  (Prada
Moroni  \& Straniero  2002,  Straniero  et al.  2003).  We treated  the
extra$-$mixing as a time-dependent  diffusion process, assuming that the
mixing   velocities  decay   exponentially   beyond  each   convective
boundary. The diffusion coefficient is  given by $D_{\rm EM}=D_0 \exp (-2 z
/f H_P)$, where  $H_P$ is the pressure scale  height at the convective
boundary, $D_0$ is the diffusion coefficient of unstable regions close
to the convective boundary, and $z$ is the geometric distance from the
edge of the convective boundary  (Herwig et al.  1997, 2000). The free
parameter   $f$  describes  the   efficiency  of   the  extra$-$mixing
process. It can take values  as high as $f\sim 1.0$, for over-adiabatic
convective   envelopes   of  DA   white   dwarfs   (Freytag  et.    al
1996). However, for deep envelope  and core convection $f$ is expected
to   be    considerably   smaller    because   the   ratio    of   the
Brunt$-$V\"ais\"al\"a  timescales  of the  stable  to unstable  layers
decreases with depth. We assume $f = 0.016$, which
accounts for the  location of the upper envelope  on the main sequence
for a large sample of clusters and associations (Schaller et al. 1992,
Herwig  et al.  1997,  2000).  The same value accounts  for the  intershell
abundances of hydrogen$-$deficient  post$-$AGB remnants (see Herwig et
al. 1997; Herwig 2000; Mazzitelli  et al. 1999). Finally, for the mass
range considered in this work, the mass of the outer convection zone on
the tip  of the RGB  only increases by  $\sim1.5 \%$ if we  change the
parameter $f$  by a  factor of two.   The breathing  pulse instability
occurring towards  the end  of the core  helium burning is
attributed  to the  adopted algorithm  rather than  to the  physics of
convection  and therefore  were  suppressed in  our computations  (see
Straniero et al.  2003 for a detailed discussion).

During  the thermally  pulsing  AGB phase,  extra$-$mixing episodes  were
disregarded. This  may prevent the  third dredge$-$up to occur  in low
mass stars. In particular, a strong reduction of extra$-$mixing episodes
at the base of the  pulse-driven convection zone seems to be supported
by  simulations  of  the  $s-$process abundance  patterns  (Lugaro  et
al.  2003) and by  observational  inferences of  the
IFMR (Salaris et al. 2009).  As a consequence,
we  expect  that the  mass  of the  hydrogen  free  core to  gradually
increase  as   evolution  proceeds  during  this   stage.  Kalirai  et
al. (2014)  estimated a growth  of the core  mass on the TP-AGB  to be
between 10\% and  30\% for initial masses  between $1.6 M_{\odot}$
and  $2.0  M_{\odot}$,  that  decreases  steadily to  $\sim  10$\%  at
$M_{\rm ZAMS}=3.4 M_{\odot}$. However, third-dredge--up episodes are not 
completely suppressed in our models. In particular, the  two more  
massive sequences  for $Z=0.0001$,  those  with initial
masses 2.0 and 2.5 $M_{\odot}$ do experience dredge--up episodes during
the TP-AGB. We will comment on the effect of third dredge--up episodes on our 
determination of the final masses in section \ref{AGB}.
In addition, the suppression of dredge--up episodes in our models prevents the 
formation of carbon rich stars, i.e, the C/O ratio remains below unity during all the TP-AGB stage.

Mass  loss was  considered during  red  giant branch  and core  helium
burning phases  following Schr\"oeder \& Cuntz (2005).  This mass$-$loss
rate  is based  on the Reimers  formula  $\dot{M} =  \eta L_*  R_* /  M_*$
(Reimers  1975),  but  it contains  two  additional factors,  one
depending  on the  surface temperature  and the  other on  the surface
gravity of  the star.  This improves the  agreement with  the observed
mass$-$loss  rates for  different types  of stars,  without the  need to
adjust the fitting parameter $\eta$ (Shr\"oeder \& Cuntz 2005). We set
$\eta=8 \times  10^{-14} M_{\odot} {\rm yr^{-1}}$,  that satisfies the
well constrained RGB mass loss of globular cluster stars.

During  the AGB  and TP-AGB  phases,  we employ  the prescriptions  of
Groenewegen et al.  (2009) and  Vassiliadis \& Wood (1993) for low and
high metallicity sequences, respectively.
The mass loss prescription presented by Vassiliadis \& Wood (1993) for
the  TP-AGB  phase  is  pulsation  dependent,  with  an  exponentially
increasing rate for  periods less than $\sim 500$  days and a constant
superwind phase for periods beyond  $\sim 500$ days, corresponding to a
radiation$-$pressure  driven wind.  This  prescription, or  an updated
version, is usually  adopted to model the mass  loss during the TP-AGB
stage   in   evolutionary   computations   (e.g. Renedo   et   al.    2010, 
Karakas \&  Lattanzio 2007,  Karakas 2010,
Marigo \& Girardi 2007). In particular, Doherty et al. (2014) employed
the  mass$-$loss  rate prescription  of  Vassiliadis \&  Wood (1993)  to
determine the  IFMR of  intermediate mass stars
with initial masses between 5.0 and 10.0 $M_{\odot}$ for metallicities
spanning  the  range of $Z=0.02-0.0001$. Groenewegen  et  al.
(2009) presented  a mass loss  prescription based on data  of resolved
stars  on  the Small  and  Large  Magellanic Clouds obtained  with Spitzer
IRS that  represents low  metallicity populations.  They  found that,
even when  the Vassiliadis  \& Wood (1993)  models described  the data
better, there are some deficiencies, in particular the maximum adopted
mass$-$loss rate.  The assumption of  typical velocities
and dust-to-gas ratios  of galactic stars may lead  to an overestimate
of the  mass$-$loss rate. Therefore  we employ the mass  loss prescription of
Groenewegen et al.   (2009) for low metallicity sequences as it is
better   calibrated  in   this  metallicity   range.    
For metallicities
larger than $Z=0.01$, mass$-$loss rate from Groenewegen et al. (2009) are
systematically lower  than those from  Vassiliadis \& Wood  (1993) and
the well  known superwind  phase at the  end of the  thermally pulsing
stage is  rarely reached  even after the  star moves  significantly to
redder  effective  temperatures.   For   this  reason  we  employ  the
prescription of Groenewegen et  al.  (2009) for evolutionary sequences
characterized  by   initial  metallicities  $Z\leq   0.008$;  and  the
Vassiliadis \& Wood (1993) mass$-$loss rates for sequences with initial
metallicities $Z\geq 0.01$, that  better reproduce mass loss rates for
high $Z$.

\subsection{White dwarf evolution}  
  
During the  white dwarf evolution,  we consider the  distinct physical
process  that   modify  the  chemical   abundances  distribution.  For
instance, element diffusion strongly modifies the chemical composition
profile throughout the outer layers. Indeed, our sequences developed a
pure  hydrogen   envelope  with  increasing   thickness  as  evolution
proceeds.  Our treatment of time$-$dependent diffusion is based on the
multicomponent gas treatment presented in Burgers (1969).  We consider
gravitational  settling and  thermal and  chemical diffusion  of $¹$H,
$^3$He, $^4$He, $^{12}$C, $^{13}$C,  $^{14}$N and $^{16}$O (Althaus et
al.  2003).  To  account for convection in our models we  adopted the mixing
length theory, in its ML2 flavor, with the free parameter $\alpha = 1$
(Tassoul et  al.  1990).  The metal  mass fraction in  the envelope is
not  fixed,  but  it   is  computed  consistently  according  to  the
predictions  of element diffusion. To account for the different envelope compositions, 
we considered radiative opacity tables from OPAL for arbitrary metallicities. 
For effective temperatures less than $10\, 000$ K we include the effects 
of molecular opacity from the computations of Marigo \& Aringer (2009). In
addition,  we considered  the chemical  rehomogenization of  the inner
carbon$-$oxygen  profile  induced  by Rayleigh$-$Taylor  instabilities
following  Salaris et  al.  (1997).  For effective  temperatures below
$10\, 000$  K, outer boundary  conditions are derived  from non$-$gray
model atmospheres (Rohrmann et al.  2012).
The equation of state is that  of Segretain et al. (1994) for the high
density regime, which accounts for all the important contributions for
both the liquid and solid phases (Althaus et al. 2007), complemented at
the  low-density regime  with an  updated version  of the  equation of
state of Magni \& Mazitelli (1979). 

Finally,   cool  white   dwarf   stars  are   expected  to   undergo
crystallization as  a result of  strong Coulomb interactions  in their
dense  interiors (van  Horn  1968).  Crystallization  occurs when  the
energy  of the Coulomb  interaction between  neighboring ions  is much
larger than their  thermal energy.  This process leads to the release of
latent heat and of gravitational energy associated with changes in the
chemical composition of carbon$-$oxygen profile (Garc\'{i}a$-$Berro et
al. 1988ab). These additional energy sources associated with
crystallization were included self-consistently, modifying  the luminosity
equation to account for both the local contribution of energy released
from the core  chemical redistribution and the latent  heat, the later
being of the order of $0.77k_{B}T$ per ion.  To asses the enhancement
of oxygen in the crystallized core we used the phase diagram presented
in Horowitz et al. (2010) -- see also Hugoto et al. (2012) and Schneider 
et al. (2012) --,  consistent with the observations of Winget
et  al. (2009;  2010) of  the white  dwarf luminosity  function  of the
globular cluster  NGC 6397. Horowitz et al.  (2010) employed molecular
dynamic simulations  considering liquid  and solid phases in mixed composition,  allowing a
better determination of the melting temperature and the composition of
the liquid and  solid phases.  They predicted an  azeotopic type phase
diagram,  and  a melting  temperature  lower  than  that predicted  by
Segretain  \& Chabrier  (1993).   Note that  the  results obtained  by
Horowitz et al. (2010) are in good agreement with the results of Medin
\& Cumming (2010), based on a semi-analytic method.

Althaus et  al (2012) presented an  study of the influence  of the 
Horowitz et  al.  (2010) phase diagram on  the evolutionary properties
of white  dwarf stars, specifically  on the cooling times.  They found
that  the amount  of mass  redistributed  in the  phase separation  is
smaller  for  Horowitz et  al.   (2010)  phase  diagram than  for  the
Segretain \& Chabrier (1993) one,  leading to a smaller energy release
from carbon$-$oxygen  differentiation.  Also, the  composition changes
are less  sensitive to  the initial chemical  profile, so  the cooling
delay   is   less  affected   by  the   uncertainties   in  the
carbon$-$oxygen  initial abundances.  Romero  et al.  (2013) study  the
effects  of the two  crystallization phase  diagrams on  the pulsation
spectrum of  variable DA  white dwarf stars  with stellar  mass $\geq
0.72  M_{\odot}$.  They  showed   that,  for  a  given  stellar  mass,
computations employing the Segretain  \& Chabrier (1993) phase diagram
was  $\sim 1000$  K  higher than  that  predicted by  the Horowitz  et
al. (2010) formulation. From their asteroseismological fits they found
that the phase diagram presented in Horowitz et al.  (2010) is the one
that  better  represents  the  crystallization process  in  the  dense
interiors of white dwarf stars.

The LPCODE has  been tested against other white  dwarf evolution code,
and the  uncertainties in  the white dwarf  cooling ages  arising from
different  numerical implementations  of  stellar evolution  equations
were found to be below 2\% (Salaris et al. 2013).

\section{Model Grid and the pre-white dwarf evolution}
\label{parameters}

\subsection{Model Grid}
\label{model-grid}
              
The  current grid  of  evolutionary models  is  composed by  a set  of
sequences with initial mass between $0.8M_{\odot}$ and $3M_{\odot}$ at
the  ZAMS,  considering  also  different metallicity  values  for  the
progenitor star between $Z=0.0001$ to  $0.04$.  Note that in this mass
range,  all sequences end  their lives  as carbon$-$oxygen  core white
dwarf stars.  The initial helium content for the starting model at the
ZAMS was provided by the  relation $Y=0.245 + 2Z$, were $Z$ represents
the  initial   metallicity.  Our  results  are   summarized  in  Table
\ref{tabla-masa}, where we  list the initial and final  masses for all
metallicity values considered in this work.  For each sequence we also
list the progenitor  age, i.e., the age from the ZAMS  to the point of
maximum effective temperature before  entering the white dwarf cooling
curve ($t_{\rm  prog}$).  The  sequences with $Z=0.01$  and $Z=0.001$,
except   for  the   sequences  with   initial  mass   of   $0.92$  and
$1.1M_{\odot}$  for the  latter  $Z$,  are based on those  from  Renedo et  al.
(2010).  Since  these  authors  employ  the Segretain  et  al.  (1993)
formulation of the phase diagram for crystallization, we recalculated
the  final  evolution  of  the  white dwarf  cooling  considering  the
Horowitz et  al.  (2010)  prescription.  The remaining  sequences were
computed   specifically   for   this   work.    Additional   sequences
characterized by high and very low metallicity values, those with $Z =
0.05$ and $Z=0.00001$, were computed  to asses the dependence of the final mass with 
metallicity,  but  also of  other parameters  as the
amount of mass  lost and the size of the helium  core at the beginning
of the central helium burning stage (see section \ref{dependence}).

\begin{table*}
\centering
\caption{Initial and final stellar masses (in solar units) and progenitor ages (in Gyr) for the metallicity values considered in Figure \ref{Mi-Mf}. Values corresponding to Z=0.01 and most to Z=0.001 are extracted from Renedo et al. (2010) (see text for details). }
\begin{tabular}{ccccccccccccccc}
\hline\hline
$M_{\rm ZAMS}$  & \multicolumn{2}{c}{$Z=0.0001$} & \multicolumn{2}{c}{$Z=0.001$} & \multicolumn{2}{c}{$Z=0.004$} & \multicolumn{2}{c}{$Z=0.008$} & \multicolumn{2}{c}{$Z=0.01$} & \multicolumn{2}{c}{$Z=0.02$} & \multicolumn{2}{c}{$Z=0.04$} \\
    & $M_{\rm WD}$ & $t_{prog}$ & $M_{\rm WD}$ & $t_{prog}$ & $M_{\rm WD}$ & $t_{prog}$ & $M_{\rm WD}$ & $t_{prog}$  & $M_{\rm WD}$ & $t_{prog}$  & $M_{\rm WD}$ & $t_{prog}$ & $M_{\rm WD}$ & $t_{prog}$ \\
\hline
0.80 & 0.519    & 12.742  & $\cdots$ &$\cdots$ & $\cdots$ & $\cdots$& $\cdots$ &$\cdots$ &$\cdots$ &$\cdots$ & $\cdots$ &$\cdots$ & $\cdots$ & $\cdots$ \\
0.85 & 0.534    & 10.327  & 0.505    & 13.322  & $\cdots$ & $\cdots$& $\cdots$ &$\cdots$ &$\cdots$ &$\cdots$ & $\cdots$ &$\cdots$ & $\cdots$ & $\cdots$ \\
0.90 & 0.550    &  8.421  & $\cdots$ & $\cdots$& 0.503    & 11.898  & 0.489    & 15.767  &$\cdots$ &$\cdots$ & $\cdots$ &$\cdots$ & $\cdots$ & $\cdots$ \\
0.92 & $\cdots$ & $\cdots$& 0.536    &  9.789  & 0.515    & 10.978  & $\cdots$ &$\cdots$ &$\cdots$ &$\cdots$ & $\cdots$ &$\cdots$ & $\cdots$ & $\cdots$\\
0.94 & $\cdots$ & $\cdots$& $\cdots$ &$\cdots$ & $\cdots$ & $\cdots$& 0.508    & 13.037  &$\cdots$ &$\cdots$ & $\cdots$ &$\cdots$ & $\cdots$ & $\cdots$  \\
0.95 & 0.561    &  6.998  & $\cdots$ &$\cdots$ & 0.524    &  9.772  & $\cdots$ &$\cdots$ & 0.493   & 13.350  & 0.488    & 15.160  & $\cdots$ & $\cdots$\\
0.98 & $\cdots$ &$\cdots$ & $\cdots$ &$\cdots$ & 0.533    &  8.736  & 0.516    & 11.679  &$\cdots$ &$\cdots$ & $\cdots$ &$\cdots$ & $\cdots$ & $\cdots$ \\
1.00 & 0.569    &  5.871  & 0.553    & 7.442   & 0.537    &  8.076  & 0.524    & 10.788  & 0.525   & 11.117  & 0.511    & 12.512  & 0.511    & 13.327  \\
1.05 & $\cdots$ & $\cdots$& $\cdots$ & $\cdots$& $\cdots$ &$\cdots$ & 0.531    &  8.053  &$\cdots$ &$\cdots$ & $\cdots$ &$\cdots$ & $\cdots$ &$\cdots$ \\
1.10 & $\cdots$ &$\cdots$ & 0.566    & 5.274   & 0.547    & 5.800   & 0.538    &  7.339  &$\cdots$ &$\cdots$ & $\cdots$ &$\cdots$ & $\cdots$ & $\cdots$ \\
1.20 & $\cdots$ &$\cdots$ & $\cdots$ & $\cdots$& 0.553    &  4.310  & 0.552    &  5.047  &$\cdots$ &$\cdots$ & $\cdots$ &$\cdots$ & $\cdots$ & $\cdots$ \\ 
1.25 & 0.621    & 2.826   & 0.593    &  3.449  & $\cdots$ &$\cdots$ & $\cdots$ &$\cdots$ &$\cdots$ &$\cdots$ & $\cdots$ &$\cdots$ & $\cdots$ & $\cdots$ \\
1.50 & 0.669    & 1.593   & 0.627    &  1.897  & 0.599    &  2.080  & 0.588    &  2.382  & 0.570   &  2.700  & $\cdots$ &$\cdots$ & $\cdots$ &$\cdots$  \\
1.75 & 0.708    & 1.038   & 0.660    &  1.324  & $\cdots$ &$\cdots$ & $\cdots$ &$\cdots$ & 0.593   &  1.699  & $\cdots$ &$\cdots$ & $\cdots$ & $\cdots$\\ 
2.00 & 0.737    & 0.752   & 0.693    &  0.947  & 0.665    &  1.074  & 0.638    &  1.074  & 0.609   &  1.211  & 0.591    &  1.286  & 0.566    & 1.368 \\
2.25 & $\cdots$ & $\cdots$& $\cdots$ & $\cdots$& 0.691    &  0.788  & $\cdots$ &$\cdots$ & 0.632   &  0.989  & $\cdots$ &$\cdots$ & $\cdots$ &$\cdots$  \\
2.50 & 0.826    & 0.423   & $\cdots$ & $\cdots$& 0.730    &  0.579  & $\cdots$ &$\cdots$ & 0.660   &  0.742  & $\cdots$ &$\cdots$ & $\cdots$ & $\cdots$ \\
3.00 & 0.875    & 0.274   & 0.864    & 0.326   & 0.817    &  0.352  & $\cdots$ &$\cdots$ & 0.705   &  0.443  & $\cdots$ &$\cdots$ & 0.656*   & 0.461 \\
\hline
\label{tabla-masa}
\end{tabular}\\
{\footnotesize{*Mass of the degenerate core at the last thermal pulse.}}
\end{table*}

\begin{figure}
\begin{center}\includegraphics[clip,width=230pt]{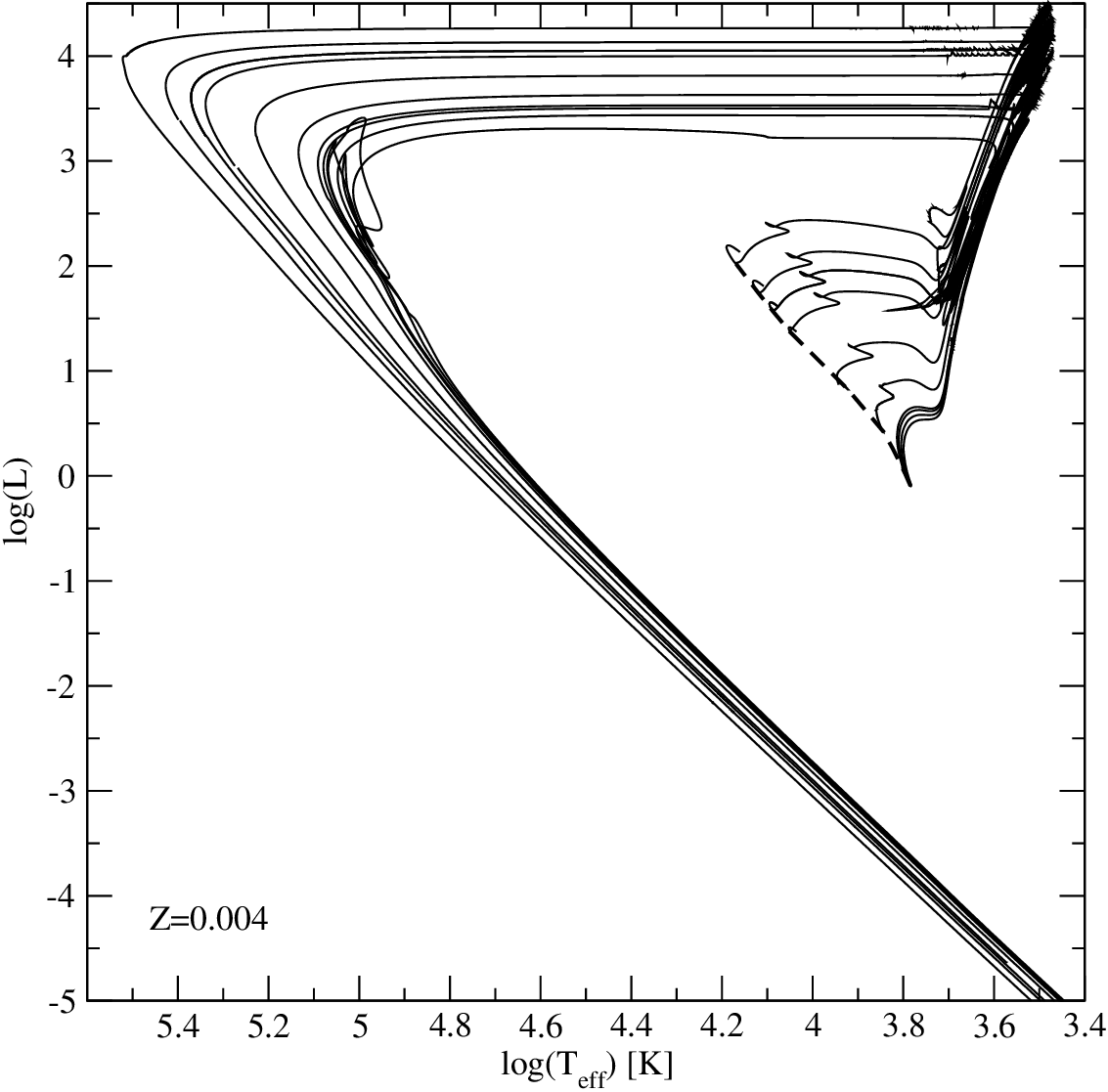}
\caption{Hertzprung$-$Russell  diagram of  our  evolutionary sequences 
for  $Z=0.004$.  From bottom  to  top: evolution  of the  
0.92$M_{\odot}$, 0.95$M_{\odot}$,  0.98$M_{\odot}$,  1.0$M_{\odot}$, 1.2$M_{\odot}$, 
1.5$M_{\odot}$, 2.0$M_{\odot}$, 2.25$M_{\odot}$, 2.5$M_{\odot}$ and 3.0$M_{\odot}$ 
model stars.  The dashed line represents the location of the Zero Age Main 
Sequence. Note that the sequences with an initial mass of 0.92$M_{\odot}$ and 
0.95$M_{\odot}$ experience an hydrogen sub-flash at high effective 
temperatures, before entering the cooling sequence.}
\label{tracks-0.008}
\end{center}
\end{figure}

The evolution  in the  Hertzprung$-$Russell diagram 
frm the  ZAMS  to  advanced  stages of  white  dwarf  evolution  for
$Z=0.004$  is shown  in Figure  \ref{tracks-0.008} for  sequences with
initial masses  larger than $0.90M_{\odot}$.  Typically, sequence with
initial mass  lower than  $\sim 0.95M_{\odot}$ experience  an hydrogen
sub-flash   before  entering  its   final  cooling   track.   Hydrogen
sub-flashes are a consequence of  an increase in the luminosity due to
shell  hydrogen  burning at  high  effective  temperatures during  the
evolution towards  the white dwarf  cooling curve.  For  $Z=0.004$ the
hydrogen  sub-flashes  are  present   for  sequences  with  $0.9$  and
$0.92M_{\odot}$.  In addition, sequences  characterized by  the lowest
initial stellar mass on the ZAMS usually do not go through the AGB and
thermally pulsing  AGB stages ($0.90M_{\odot}$  for $Z=0.004$).  These
sequences spend the core helium burning  stage at the blue edge of the
Horizontal Branch phase  and move directly to the  white dwarf cooling
curve without experiencing the high luminosity phase.

\begin{figure}
\centering
\includegraphics[clip,width=0.45\textwidth, angle=0]{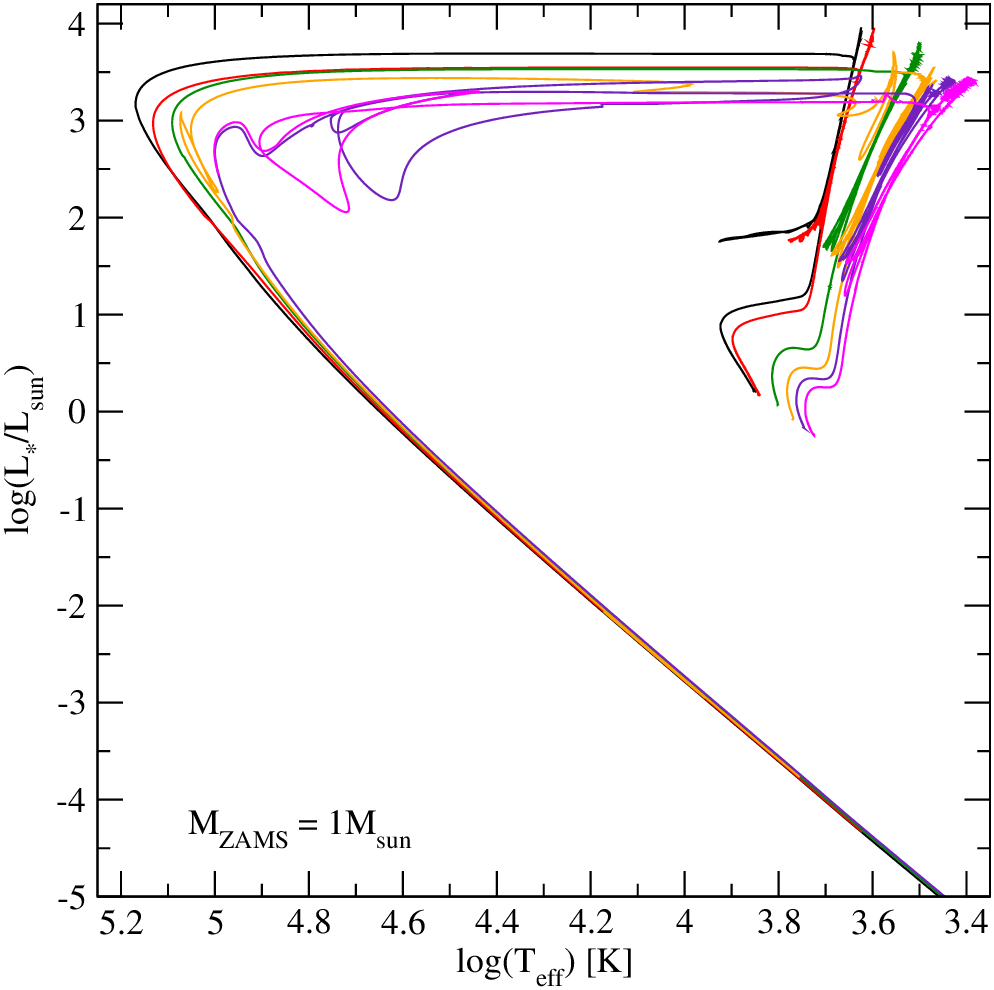} 
\caption{Hertzprung$-$Russell  diagram of  our  evolutionary sequences with initial stellar mass of 1$M_{\odot}$ with metallicities of $Z=0.00001, 0.0001, 0.004, 0.01, 0.02$ and $0.04$ from top to bottom. }
\label{H-R-Z}
\end{figure}

In Fig.  \ref{H-R-Z} we  show the evolutionary sequences characterized
by  a  $1M_{\odot}$ at  the  ZAMS  and  different initial  metallicity
values.   When  metallicity  increases,   the  position  of  the  main
sequences moves towards lower luminosities and effective temperatures,
since the atmospheres of more metallic stars are more opaque, and then
the larger opacities make the models move towards redder temperatures.
On the other  hand less metallic models are  more luminous because the
atmospheres are more  transparent.  This is also true  when we inspect
the AGB and TP-AGB stages; less metallic models are dislocated to the
blue  because  of the  lower  opacities.   As  we mention  in  section
\ref{pre-wd} the mass--loss  rate is metallicity dependent (Vasiliadis
\&  Wood  1993; Groenewegen  et  al.  2009),  then the  less  metallic
sequences have less efficient mass  loss episodes, leading to a larger
number of thermal pulses.

In the present work, the  hydrogen content of the sequences correspond
to the  value predicted  by standard stellar  evolution and  should be
considered as  upper limits for  the maximum hydrogen content  left in
the  white   dwarf  resulting  for   the  evolution  of   single  star
progenitors.  As we  will show  in section  \ref{H-env}, the  amount of
hydrogen left  on a white dwarf  envelope depends on  stellar mass and
metallicity (see e.g. Renedo  et al. 2010, Romero  et al. 2012).  The amount of
hydrogen  left  in  a white  dwarf  star  is  an important  factor  to
determine the cooling  ages, since hydrogen has a  larger opacity than
helium.  Cooling  ages  can  be  reduced by  10\%  for  thin  hydrogen
envelopes  ($10^{-10} M_*$)  at $\log(L/L_{\odot})=-5.5$  (Catal\'an et
al. 2008a) in comparison with thick ones. In addition, a thinner hydrogen 
envelope can be mixed with
the underlying  helium layer  when the outer  convective zone  at the
white dwarf  atmosphere reaches  the hydrogen-helium interface  at low
effective temperatures,  turning the  hydrogen atmosphere star  into a
helium  atmosphere   white  dwarf  (Tremblay \& Bergeron 2008; 
Romero  et  al.   2013;  Kurtz  et
al. 2013). Evidence from asteroseismology (Castanheira \& Kepler 2009;
Romero et al.  2012, 2013) showed that the mass  of the hydrogen layer
 spreads  over $10^{-4}-10^{-10}M_*$  with  a mean
value around $10^{-6}M_*$, two orders  of magnitude smaller than the white dwarf
models in the literature employed in evolutionary studies of this type
of star.

\subsection{The core mass at helium ignition}
\label{dependence}

\begin{figure}
\begin{center}
\includegraphics[clip,width=0.48\textwidth]{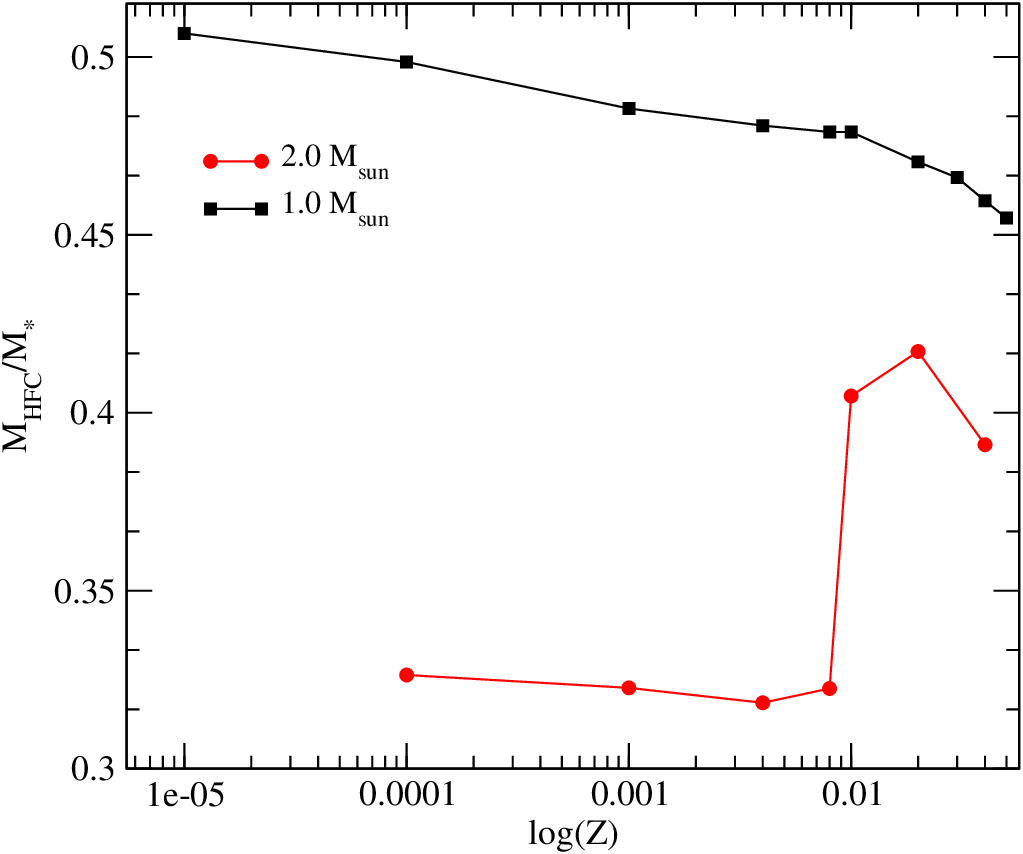}
\caption{Mass of the hydrogen free core at the beginning of the central helium burning stage as a function of metallicity, for sequences 
with $1M_{\odot}$ (squares) and $2M_{\odot}$ (circles) at the ZAMS. Note that the larger the mass at the ZAMS, the smaller the mass of the hydrogen free core.}
\label{Mi-Mf-1}
\end{center}
\end{figure}

In this  section we analyze the  impact of metallicity on  the mass of
the helium core at the  beginning of the central helium burning stage,
and also  on the total  mass lost in  the different giant  phases.  In
Table  \ref{tabla-Z} we  list the  final mass  values along  with some
other relevant  parameters for all  sequences with initial mass  of $1
M_{\odot}$ and  $2 M_{\odot}$.  We computed additional sequences, apart 
from those listed in Table \ref{tabla-masa} with
$Z=0.00001$, $Z=0.03$ and  $Z=0.05$ for $M_{\rm ZAMS}=1M_{\odot}$, and
$Z=0.03$ for  $M_{\rm ZAMS}=2M_{\odot}$.   The second column  of Table
\ref{tabla-Z} lists  the mass  of the helium  core before the  onset of
helium burning,  i.e., the hydrogen free  core mass at the  end of the
RGB phase.  For  all sequences with a $1M_{\odot}$  progenitor and the
sequences  with a  progenitor star  with $2M_{\odot}$  and metallicity
$Z\geq 0.01$, this  mass is the mass of the helium  core at the maximum
of  the helium  luminosity during  the  helium flash.   Note that  for
sequences with initial  mass $1M_{\odot}$, the mass of  the helium core
increases with decreasing  metallicity approximately $0.052 M_{\odot}$
in the range considered in this work.  This can be seen also from Fig.
\ref{Mi-Mf-1}  were  we  plot the  mass  of  the  helium core  at  the
beginning  of  the central  helium  burning  stage  as a  function  of
metallicity for  sequences with  initial stellar mass  of $1M_{\odot}$
(squares) and $2M_{\odot}$ (circles).  

\begin{table*}
\centering
\caption{Different parameters as a function of the metallicity of the progenitor star with 
initial mass $1M_{\odot}$ (squares) and $2M_{\odot}$ (circles) at the ZAMS. The columns list the metallicity value, 
the mass of the hydrogen free core at the helium flash or at the beginning of the helium 
central burning stage, the mass lost during the Red Giant Branch and Asymptotic Giant Branch 
phases and the final mass at the white dwarf cooling sequence. All masses are in solar units. }
\begin{tabular}{lcccccccc}
\hline\hline
 & \multicolumn{4}{c}{$M_{\rm ZAMS}=1M_{\odot}$} & \multicolumn{4}{c}{$M_{\rm ZAMS}=2M_{\odot}$}\\
$Z$  &  $M_{\rm HFC}$ & $\Delta M({\rm RGB})$ & $\Delta M({\rm AGB})$ & $M_{\rm WD}$ &  $M_{\rm HFC}$ & $\Delta M({\rm RGB})$ & $\Delta M({\rm AGB})$ & $M_{\rm WD}$\\
\hline
0.00001 &  0.5066  &    0.1068  &   0.3280 & 0.564  & $\cdots$ & $\cdots$ & $\cdots$ & $\cdots$ \\
0.0001  &  0.4986  &    0.1374  &   0.2960 & 0.562  & 0.3263  &  0.0001  &  1.2006 & 0.738\\
0.001   &  0.4855  &    0.1782  &   0.2535 & 0.553  & 0.3227  &  0.0021  &  1.2725 & 0.693 \\
0.004   &  0.4807  &    0.2342  &   0.2168 & 0.537  & 0.3185  &  0.0023  &  1.3195 & 0.665 \\
0.008   &  0.4789  &    0.2834  &   0.1818 & 0.524  & 0.3225  &  0.0028  &  1.3492 & 0.638  \\
0.01    &  0.4789  &    0.3218  &   0.1410 & 0.525  & 0.4047  &  0.0172  &  1.3565 & 0.609 \\  
0.02    &  0.4705  &    0.3628  &   0.1169 & 0.511  & 0.4172  &  0.0243  &  1.3777 & 0.591 \\
0.03    &  0.4661  &    0.3870  &   0.0883 & 0.511  & $\cdots$ & $\cdots$ & $\cdots$ & $\cdots$\\   
0.04    &  0.4596  &    0.3912  &   0.0887 & 0.511  & 0.3910  &  0.0149  &  1.4126 & 0.566 \\ 
0.05    &  0.4547  &    0.3747  &   0.1004 & 0.515  & $\cdots$ & $\cdots$ & $\cdots$ & $\cdots$\\
\hline\hline
\label{tabla-Z}
\end{tabular}
\end{table*}

For sequences with progenitor mass of $2M_{\odot}$ we have two groups,
those  sequences  with  initial  metallicity  below  $Z=0.01$  do  not
experience an  helium ignition on a degenerate core. As a  consequence the mass  of the helium
core at the  beginning of the central helium  burning is significantly
lower, around  $0.32-0.33M_{\odot}$. On the other  hand, sequences with
initial  metallicities ranging from  0.01 to  0.04 ignite  the central
helium in a degenerate state.  Although the mass of the helium core at
this point  is lower than that  for the $1M_{\odot}$  sequences, it is
higher compared  to the metal poor  cases. This change  in the central
helium core  mass is  clear in Fig.  \ref{Mi-Mf-1}. In  particular, we
note  that the  helium  core mass  is  maximum for  $Z=0.02$ and  then
decreases for lower and higher metallicities. Also, the maximum helium
luminosity  at  the  moment of  the  helium  flash  is 3-4  orders  of
magnitude larger for the sequences with $Z=0.02$.
From stellar  evolutionary theory we know  that the size  of the final
hydrogen free core is mainly  controlled by two process: mass--loss and
overshooting.  Mass--loss rates decreases  with decreasing metallicity,
allowing  the core to  increase its  size given  the larger  amount of
burning  material. On  the other  hand,  even though  mass--loss  rates
increases  with metallicity,  the  influence of  overshooting is  more
important for high  $Z$, also allowing the core  to increase its size.
For $Z$ as high as 0.05 the overshooting influence could
be overcoming the influence of mass loss.

Also listed in Table \ref{tabla-Z}  are the total mass lost during the
RGB  and AGB  phases.  For  the sequences  with a  progenitor  star of
$1M_{\odot}$ we see  that the mass lost during  the RGB increases with
increasing metallicity,  leaving a smaller  amount of mass to  be lost
during the AGB phase.  In fact,  sequences with the highest values of Z
do  not  experience   any  thermal  pulses  and  the   growth  in  the
carbon--oxygen core  is only due to  the evolution during  the AGB.  On
the other  hand, the  low metallicity sequences  experience up  to 4--5
thermal  pulses, extending their  stay at  the high  luminosity stage.
For the sequences with a progenitor star of $2M_{\odot}$ the mass lost
during the RGB is negligible  compared with the total mass lost during
the AGB and thermal pulsing  AGB.  Practically all mass is lost during
the late  stages of the evolution.   Because of this,  the modeling of
the mass--loss rates becomes so important during the AGB and TP-AGB.

Finally,  the  dependence  of  the  final  mass  with  the  progenitor
metallicity also  has an impact on  the inner composition  of a given
white  dwarf  star.   Since  the   final  mass  is  higher  for  lower
metallicities, low mass carbon--oxygen  white dwarf can have lower mass
progenitors at lower metallicity.   On the other hand, for lower
main sequence stars,  the mass of the helium core  at the helium flash
increases with  decreasing metallicity.   For instance, a  white dwarf
with a  stellar mass of  $\sim 0.5M_{\odot}$ resulting from  a $1M_{\odot}$
progenitor star could  have an helium core if  the initial metallicity
is lower  than $Z=0.001$,  but it will  be a carbon--oxygen  core white
dwarfs  for  higher  metallicity   values,  according  to  our  single star 
evolution computations.

\subsection{Core mass growth during the thermal pulsing AGB}
\label{AGB}

 After  helium exhaustion in  the central  core, the  star moves  to the  
AGB.  In the early stages, the helium burning shell
 increases  the  carbon--oxygen  core  mass  as it  moves  towards  the
 surface.   In our  computations, overshooting  is taken  into account
 during  the early  AGB, with  an efficiency  parameter  of $f=0.016$,
 while it is disregarded  during the following thermal pulsing stage.
A reduction of extra$-$mixing
 episodes at the base of the driving convection zone during the TP-AGB
 could prevent  the third dredge--up to occur,  specially in  low mass
 stars.  The occurrence of  the third dredge--up cause an instantaneous
 reduction of  the core  mass, dragging processed  material, specially
 carbon, to the surface (see  e.g.  Wiess \& Ferguson 2009; Karakas et
 al.   2010;  Marigo et  al.   2013).   The  efficiency of  the  third
 dredge--up    is    usually     parametrized    by    the    parameter
 $\lambda$\footnote{The efficiency  parameter $\lambda$ is  defined as
   the  fraction of the  core-mass growth  over the  interpulse period
   that is dredge--up to the  surface at the next thermal pulse}, which
 increases with each thermal pulse until a maximum value $\lambda_{\rm max}$ 
(see e.g.  Marigo et al.  2013).  The value of $\lambda_{\rm max}$  
typically  increases  with  the  stellar  mass,  while  it
   decreases  at larger metallicities.   Then, dredge--up  episodes are
   expected to be more efficient for high masses and low metallicities.
   Unfortunately,  the  efficiency  $\lambda$   is  one  of  the  most
   uncertain parameters  of TP-AGB modeling, and it  is very dependent
   on the adopted treatment of convection, mixing, and numerical codes
   (Marigo et  al. 2012; Kalirai et al. 2014).  

Extra$-$mixing
   episodes  during  the  TP-AGB   also  leads  to  the  formation  of
   carbon rich stars. Studies of AGB stellar populations in our Galaxy and the Magellanic Clouds shows that stars with initial mass $< 2M_{\odot}$ and low metallicity values do become carbon rich stars (Groenewegen et al. 1993; Stancliffe et al. 2005; Marigo \& Girardi 2007). For instance, carbon rich stars must be formed for initial masses as low as $1M_{\odot}$ in the Large Magellanic Cloud in order to reproduce the carbon stars luminosity function (Sancliffe et al. 2005).  Carbon enrichment  of the atmosphere will cause an abrupt change in the  molecular equilibrium leading to a rise of the atmospheric opacity.  This causes  an increase in the mass--loss rate, leading to  an early end of the  TP-AGB and thus a reduction of the final mass.
   Because in  our computations we  disregarded extra$-$mixing episodes,
   our  sequences in general do not  experience  dredge--up  episodes during  the
   TP-AGB  evolution,  except for  the  three  more massive  sequences, 
   corresponding to the  lowest metallicity  of our  grid.  In our models, sequences
   with  initial  masses  between  $2M_{\odot}$ and  $3M_{\odot}$  and
   $Z=0.0001$  do  experience  dredge--up  episodes during  the  TP-AGB
   evolution, with values of  $\lambda_{\rm max}$ of $\sim 0.22-0.36$, 
   increasing with stellar mass. The mass of the core for the onset of
   the  third  dredge--up   also  increases  from  $0.652M_{\odot}$  to
   $0.836M_{\odot}$ for  the same mass range. This values are $\sim 10-20\%$ 
larger in comparison to the results of Wiess \& 
Ferguson (2009) for $Z=0.0005$, adopting $f=0.016$. 
Finally, all our sequences
   that  experience dredge--up  episodes become  carbon-rich  stars. We
   found  that  it only  takes  one  dredge--up episode to get  a  carbon-rich
   surface, leading to a C/O ratio of  $\sim 8-10$ at the end of the TP-AGB.
   This values  of C/O  are much higher  than the C/O  $\sim 1.28-2.0$
   found  in Galactic  carbon stars  (Lambert et  al. 1986;  Ohnaka et
   al. 2000).  Note  that the input physics in  LPCODE is dedicated to
   model  in  detail   the  inner  structure  of  the   star  for  all
   evolutionary stages, and specially  the white dwarf stage. However,
   it does  not compute a detailed  atmosphere model, similar to  most full
   evolutionary  codes in  the literature. It is no tailored to
   reproduce the atmosphere abundances found in AGB stars.

To asses the effect of the extra$-$mixing episodes on our determinations of 
the final
mass,   we calculated  additional sequences  allowing extra$-$mixing
episodes  to occur, by considering  different values  of the
overshooting   parameter $f$.  Wiess  \&  Ferguson   (2009)  consider
$f=0.016$, and found  that the final mass was very similar  to that of the
first  thermal pulse, as a consequence  of the  enhanced dredge--up.  On the
other hand Lugaro et al. (2003) argued for a lower efficiency $f=0.008$,
to  achieve a  better  agreement with  detailed $s-$process observed abundance
patterns.  We computed sequences with initial mass of $2M_{\odot}$ and
metallicity $Z=0.01$ with  different values of  $f$ =0.016,
0.008 and  0.002. Note  that the core  mass growth during  the thermal
pulses  is by  far  larger  for sequences  with  initial masses  $\sim
2M_{\odot}$, where  the core mass at  the first thermal  pulse shows a
strong minimum (Renedo et al. 2010; Marigo et al. 2013; Kalirai et al. 2014). 
All three sequences computed with some amount of overshooting do become carbon stars after $10-11$ thermal pulses and $\sim 4$ dredge--up episodes, with C/O ratios of $\sim 1.3 - 1.5$ at the end of the TP-AGB evolution, and values of $\lambda_{\max}$ from 0.63 for $f=0.002$ to 0.94 for $f=0.016$.
Regarding the core mass growth, we  found that the final mass of the hydrogen free core is reduced
by 6.2\%,  7.6\% and 9.4\% when  $f$ takes values of  0.002, 0.008 and
0.016,  respectively. As a result, the  final mass can be overestimated  by less
than 10\% in our computations  for this metallicity.  For lower values
of $Z$,  like $Z=0.004$ and the  same initial mass, we  found a larger
reduction of the  final mass by $\sim 15\%$,  when $f=0.008$. This is
in agreement with the fact  that dredge--up episodes are more efficient
for lower metallicity values.  Karakas  et al. (2002) showed that even
for  low  metallicity  values  as $Z=0.004$,  $\lambda_{\rm  max}$  is
maximum  for  $M_{\rm  initial}  \sim 2.5M_{\odot}$  and  then  slowly
decreases for larger masses, leading  to a less efficient reduction of
the core mass due to dredge--up episodes. Then, in the worst case scenario, the uncertainties in final
masses from our treatment of overshooting episodes might be up to $\sim 15\%$ for lower metallicities.
Note that, additional uncertainties on the final core mass arise from the possible reduction of TP-AGB lifetimes when C--rich molecular opacities are considered, leading to an increase in the surface opacity and an enhanced mass--loss rate at this stage.

Finally, Kalirai  et al.  (2014) performed an  study of the  core mass
growth  covering the intermediate-mass  range of  the IFMR. 
They use data from four open clusters with nearly solar
metallicities:  NGC 6819,  NGC 7789,  Hyades and  Praesepe,  the
first  three included  in our  analysis of  the  initial-to-final mass
relation (see  section \ref{i-f}).  As a result,  they found  that, in
order  to  fit the  data,  the  dredge--up efficiency  $\lambda_{\max}$
increases  from  zero to  $\sim  0.5$  for  initial mass  from  $\sim
2M_{\odot}$ to $\sim 3M_{\odot}$  and then progressively decreases for
larger initial masses.  From our theoretical results with $Z=0.002$ we
found a core mass growth of  $\sim 10\%$ to $\sim 28\%$ for stars with
initial mass $1.6M_{\odot}$  and $2.8M_{\odot}$, decreasing steadily to
$\sim 6\%$ at $M_{\rm initial} = 3.7 M_{\odot}$, somewhat smaller than
the  results  of  Kalirai  et  al.  (2014),  but  still  in  excellent
agreement.   In  addition,  Kalirai  et  al.  (2014)  found  that  the
dominant process  governing the growth of the  core is largely the
stellar  wind, while  the  third dredge--up,  although not  negligible,
plays  a  secondary  role.   They   also  found  that  the  mass  loss
prescription  that better  fits  the semi-empirical  data  is that  of
Vasiliadis \& Wood (1993), the same mass loss prescription employed in
our work for solar metallicity.

\section{Initial--to--final mass relation}
\label{i-f}

\begin{figure*}
\begin{center}
\includegraphics[clip,width=0.70\textwidth]{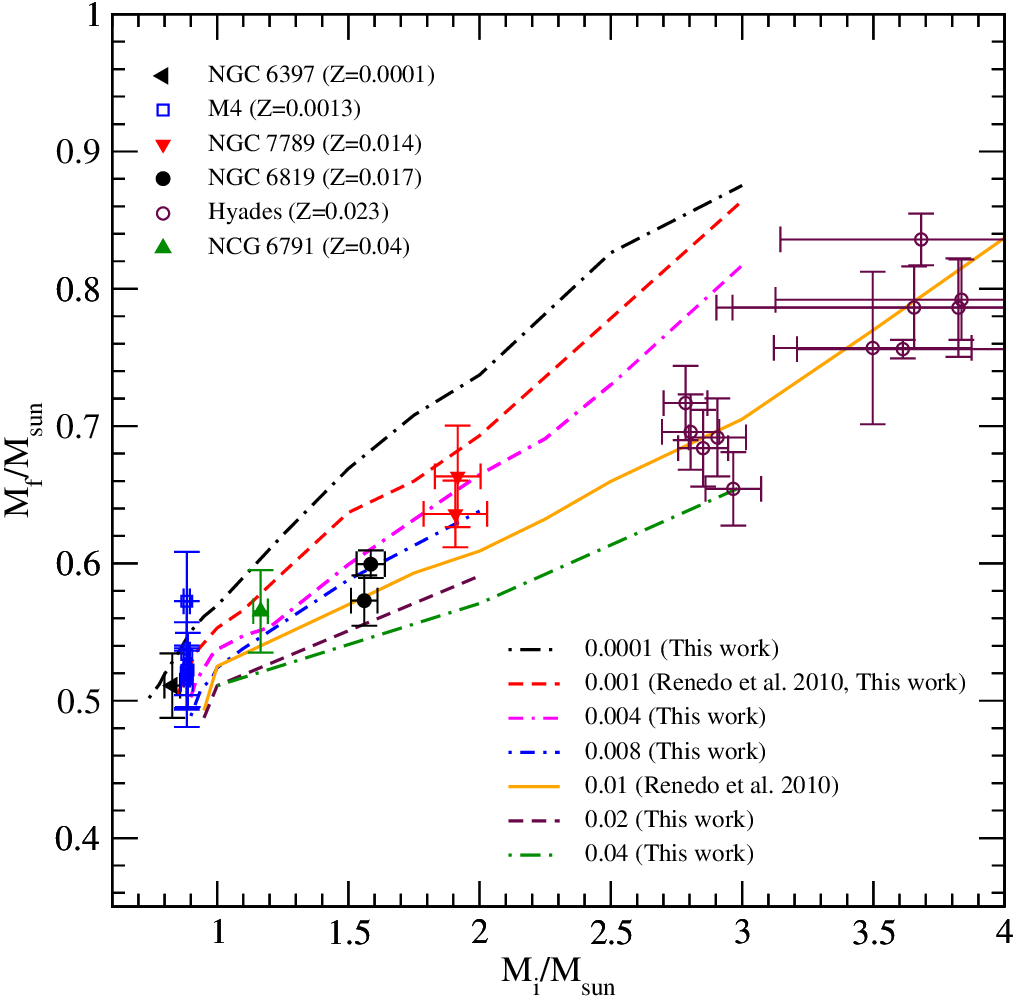}
\caption{Theoretical IFMR for different metallicity 
values from $Z=0.0001$ to $Z=0.04$ and initial mass  $\leq 3M_{\odot}$. Values 
for $Z=0.01$ and initial mass larger than $3M_{\odot}$ are taken from Renedo et al. (2010).  
All computations were made by using the LPCODE evolutionary code. We also 
include some observational data for globular and old open clusters -- see 
text for details.
(A color version of this figure is available in the on-line journal.) }
\label{Mi-Mf}
\end{center}
\end{figure*}

In this section  we show our results for the  theoretical IFMR and its
dependence with  the different metallicity.  Given our  model grid, we
focus on the low and intermediate mass domain of the  IFMR, and consider sequences
with an  initial mass $\leq  3M_{\odot}$.  This accounts  for globular
clusters and old open clusters, having main sequence stars within this
mass range.  For  young open clusters with ages of  the order of $\sim 100$  Myr, low mass  stars have  no time  to evolve  out of  the main
sequences  yet.  The  initial mass  ($M_{\rm ZAMS}$),  the  final mass
($M_{\rm  WD}$)  and the  progenitor  lifetime  for each  evolutionary
sequences   for   all   the   metallicities  are   listed   in   Table
\ref{tabla-masa}.

We consider  the final  mass as that  of the  white dwarf star  in the
cooling curve,  where the mass--loss  rate becomes negligible,  not the
mass of the hydrogen free core in  the first thermal pulse. As show in
section  \ref{AGB}  and by  Kalirai  et al.  (2014)  the  mass of  the
hydrogen free  core increases  during the TP-AGB  as much as  30\% for
$M_{\rm  initial}\sim  2.0M_{\odot}$.   Then,  the IFMR  for  a  fixed
metallicity by  the end of the thermally pulsing  AGB becomes markedly
different from that  determined by the mass of  the hydrogen free core
at the first thermal pulse (Althaus et al. 2010).  For instance, if we
take  a white  dwarf  with a  stellar  mass of  $0.63 M_{\odot}$,  the
initial mass  would be  overestimated by $\sim  0.75 M_{\odot}$  if we
were to consider an IFMR at the first thermal-pulse, instead of the real one, at
advanced stages  in the thermally pulsing AGB.   For the carbon--oxygen
composition  expected  in  the core  of  a  white  dwarf, this  is  an
important issue (Althaus et al. 2010).

In Figure  \ref{Mi-Mf} we plot  the initial--to--final mass relation. Each
curve  represents  a set  of  theoretical  computations  with a  given
initial  metallicity  value. From  this
figure  we  can see  that  the  IFMR shows  a
dependence  with  metallicity:  for   a  given  initial  mass  of  the
progenitor  star,  the  mass  of  the resulting  white  dwarf  remnant
decreases when  metallicity increases.  This effect  is noticeable for
progenitor mass  above $\sim 1M_{\odot}$.   For instance, the  mass of
the resulting white dwarf evolving from a $2M_{\odot}$ progenitor star
is $\sim 21 \%$ larger  when the metallicity decreases from $Z=0.0001$
to 0.01.  We expect the difference  to be even larger for more massive
stars. For low  mass progenitor the differences are  smaller, but still
not-negligible. For  a star with  an initial mass at  $0.85 M_{\odot}$
there is an increase in the  white dwarf mass of $\sim 0.03 M_{\odot}$
when  the metallicity  decreases from  0.001  to 0.0001.  Even if the
differences in the  final mass are not very  large, the differences in
the time spent during the stages previous to the white dwarf sequence
will be, because the evolutionary time  scales for low mass stars are very
long.  This  age difference  is  important  when  we use  evolutionary
computations  to  estimate  the   age  of  stellar  populations.

\begin{table*}
\centering
\caption{Results from our fittings for the initial and final masses for the white dwarf stars corresponding to five different clusters. Effective temperature and superficial gravity values for NGC 6819, NGC 7789 and Hyades are taken from Kalirai et al. (2014), for NGC 6791 are taken from Kalirai et al (2007), for NGC 6397 are taken from Moehler et al. (2004) and for M4 are from Kalirai et al. (2009). }
\begin{tabular}{ccccccc}
\hline\hline
Cluster & star & $T_{\rm eff}$ (K) &  $\log g$ & $M_{\rm WD}/M_{\odot}$ & $t_{\rm cool}(Myrs)$ & $M_{\rm initial}/M_{\odot}$ \\  
\hline
NGC 6397 & WF4-358    & $18800\pm 340$ & $7.72\pm 0.06$ & $0.511\pm 0.023$ & $75.75\pm 8$    & $0.829\pm 0.030$\\
M4       & WD-00      & $20600\pm 600$ & $7.75\pm 0.09$ & $0.522\pm 0.027$ & $52.89\pm 8$    & $0.885\pm 0.011$\\
M4       & WD-04      & $24300\pm 500$ & $7.68\pm 0.08$ & $0.516\pm 0.028$ & $27.48\pm 3$    & $0.885\pm 0.011$\\
M4       & WD-06      & $25600\pm 500$ & $7.87\pm 0.08$ & $0.572\pm 0.036$ & $21.08\pm 5$    & $0.885\pm 0.011$\\
M4       & WD-15      & $24300\pm 600$ & $7.79\pm 0.09$ & $0.534\pm 0.039$ & $27.29\pm 3$    & $0.885\pm 0.011$\\
M4       & WD-20      & $19700\pm 600$ & $7.73\pm 0.10$ & $0.519\pm 0.038$ & $58.83\pm 17$   & $0.885\pm 0.011$\\
M4       & WD-24      & $25700\pm 500$ & $7.70\pm 0.07$ & $0.522\pm 0.018$ & $22.32\pm 3$    & $0.885\pm 0.011$\\
NGC 7789 & NGC 7798-5 & $31600\pm 200$ & $7.89\pm 0.05$ & $0.636\pm 0.024$ & $8.66\pm 0.3$   & $1.908\pm 0.051$\\
NGC 7789 & NGC 7798-8 & $25000\pm 400$ & $8.06\pm 0.07$ & $0.663\pm 0.037$ & $26.15\pm 9$    & $1.917\pm 0.051$\\
NGC 6819 & NGC 6816-6 & $21900\pm 300$ & $7.89\pm 0.04$ & $0.532\pm 0.017$ & $43.15\pm 4$    & $1.561\pm 0.050$\\
NGC 6819 & NGC 6816-7 & $16600\pm 200$ & $7.97\pm 0.04$ & $0.599\pm 0.010$ & $141.63\pm 15$  & $1.585\pm 0.054$\\
Hyades   & WD0325+096 & $14670\pm 380$ & $8.30\pm 0.05$ & $0.786\pm 0.030$ & $357.59\pm 69$  & $3.655\pm 0.692$\\ 
Hyades   & WD0406+169 & $15810\pm 290$ & $8.38\pm 0.05$ & $0.836\pm 0.019$ & $337.26\pm 43$  & $3.682\pm 0.536$\\
Hyades   & WD0421+162 & $20010\pm 320$ & $8.13\pm 0.05$ & $0.692\pm 0.028$ & $97.75\pm 23$   & $2.906\pm 0.109$\\
Hyades   & WD0425+168 & $25130\pm 380$ & $8.12\pm 0.05$ & $0.696\pm 0.027$ & $30.48\pm 6$    & $2.804\pm 0.109$\\
Hyades   & WD0431+126 & $21890\pm 350$ & $8.11\pm 0.05$ & $0.684\pm 0.028$ & $61.76\pm 13$   & $2.851\pm 0.096$\\
Hyades   & WD0437+138 & $15120\pm 360$ & $8.25\pm 0.09$ & $0.757\pm 0.055$ & $305.00\pm 59$  & $3.498\pm 0.377$\\
Hyades   & WD0438+108 & $27540\pm 400$ & $8.15\pm 0.05$ & $0.717\pm 0.028$ & $17.65\pm 5$    & $2.785\pm 0.083$\\
Hyades   & WD0348+350 & $14820\pm 350$ & $8.31\pm 0.05$ & $0.792\pm 0.029$ & $363.95\pm 73$  & $3.836\pm 0.709$\\
Hyades   & HS0400+1451 & $14620\pm 60$ & $8.25\pm 0.01$ & $0.756\pm 0.068$ & $325.17\pm 23$  & $3.612\pm 0.403$\\
Hyades   & WD0625+415 & $17610\pm 280$ & $8.07\pm 0.05$ & $0.654\pm 0.027$ & $137.70\pm21$   & $2.967\pm 0.106$\\
Hyades   & WD0637+477 & $14650\pm 590$ & $8.30\pm 0.06$ & $0.786\pm 0.036$ & $361.94\pm 111$ & $3.824\pm 0.923$\\ 
NGC 6791 & WD7        & $14800\pm 300$ & $7.91\pm 0.06$ & $0.565\pm 0.030$ & $196.66\pm 40$  & $1.166\pm 0.028$\\
\hline
\label{determina}
\end{tabular}\\
\end{table*}

For   comparison   purposes,   we   include  in   Figure   \ref{Mi-Mf}
observational data  corresponding to  open and globular  clusters with
a Main Sequence Turn Of mass in the range $\sim 0.8-4.0M_{\odot}$.  We consider the data
from two globular  clusters, M4 and NGC 6397,  and four open clusters,
NGC  6791, NGC  6819, NGC  7789 and  Hyades.  The  cluster and star
identifications and  the spectroscopic parameters for  the star sample
are  listed  in  the  first  four columns  of  Table  \ref{determina}.
Effective temperature  and superficial gravity values  were taken from
the works of Kalirai et al.  (2014) for NGC 6819, NGC 7789 and Hyades,
from Kalirai  et al. (2007) for  NGC 6791, from Moehler  et al. (2004)
for  NGC  6397  and  from   Kalirai  et  al.  (2009)  for  M4.   These
spectroscopic  parameters were  derived  from the  well know  technique
which involves fitting the Balmer lines shape of the spectrum to model
atmospheres  (e.g. Bergeron  et al.   1992).  With  $T_{\rm  eff}$ and
$\log  g$  constrained, we  estimated  the  spectroscopic mass  values
(column  5 of  Table  \ref{determina}) by  a  linear interpolation  of
carbon--oxygen  core cooling  tracks in  the  $\log g  - T_{\rm  eff}$,
computed  from full  evolutionary models. 
The cooling  ages ($t_{\rm cool}$) for the  white dwarf stars
were extracted also from our evolutionary tracks (column 6 from Table
\ref{determina}).  By subtracting $t_{\rm cool}$ from the total age of
the cluster, we obtained the  lifetime of the progenitor star ($t_{\rm
  prog}$) at the  point of highest temperature in  the post-AGB stage,
before entering  the white dwarf cooling curve.  
We assumed  that the total ages  of the clusters are  $2.5\pm 0.2$ Gyr
for NGC 6819 and $1.4\pm 0.1$  Gyr for NGC 7789 (Kalirai et al. 2001,
2008), $625\pm 50$ Myrs for  the Hyades (Perryman et al.  1998; Claver
et al.  2001),  $8.5\pm 0.5$ Gyr for NGC 6791  (Kalirai et al. 2007),
$11.7\pm 0.3$ Gyr for NGC 6397 (Hansen et al. 2013) and $11.6\pm 0.6$
Gyr  for M4  (Bedin et  al.  2009).   The initial  mass  values (last
column  of Table  \ref{determina})  have been  computed directly  from
the pre-white dwarf lifetime $t_{\rm prog}$, using the mass-lifetime relations 
from pre-white dwarf theoretical  evolutionary  models.   For  each cluster  we  take  into
account  the  metallicity value  and  consider  a  grid of  progenitor
sequences with  the closest $Z$ to  compute the initial  mass from the
progenitor age.

Note that there is  a good agreement between theoretical IFMR
 and  the semi  empirical  values. In  general, all  the
points representing  the star  sample listed in  Table \ref{determina}
are  accounted for  by the  range covered  by our  theoretical curves,
within the  metallicity range considered  by our computations. Then, the
spread observed  in the data can  be solved by considering  a spread in
the  metallicity   values.   A   more  detailed  analysis   of  Figure
\ref{Mi-Mf} shows that for stars belonging  to M4 and NGC 6397, at the
low mass limit, and to the  Hyades, for high mass values,
the  theoretical curves  with the  right metallicity  fits  the points
within the  uncertainties.  For NGC 6819, the data is better represented with somewhat lower metallicities than the listed observational value $Z=0.017$. However, asteroseismic inferences from red giants stars from this cluster suggest that metallicity might be significantly lower than solar, between $Z=0.0033$ and $Z=0.008$ (Hekker et al. 2011). 
For the  very metal rich open  cluster NGC
6791,  represented  by  a  single  object  WD7,  and  the  near  solar
metallicity  open cluster  NGC 7789,  the accordance  between  the semi
empirical data and the  theoretical IFMR is poor. 
For  NGC 7789 the metallicity  value $Z=0.014$ is  an average of
literature  values  obtained by  means  of  different photometric  and
spectroscopic  techniques   (see  Wu  et  al.    2007),  ranging  from
$Z=0.0085$  $([Fe/H]=-0.35)$ to  solar ($\sim  Z=0.019$).  Our results
would be in  better agreement with a metallicity  value closest to the
lower limit  $\sim 0.008$.  In the case  of NGC 6791 we  expect a higher
mass--loss rate  given its  high metallicity.  However the  final mass
obtained  for  WD7 is  around  $\sim  0.03M_{\odot}$  higher than  the
expected  value, indicating that  the theoretical  mass--loss  rates at
high metallicity values should be smaller than assumed, in agreement
with the  results obtained by Miglio  et al. (2012)  for this cluster.
In  summary, different  metallicty
values of the progenitor stars  leads to different final masses on the
white dwarf colling curve,  causing the observed spreading.

\subsection{Comparison with theoretical and semi-empirical IFMR}

In  this  section  we  compare  our  results  with  other  theoretical
determinations  of the IFMR  performed using  independent evolutionary
codes. In particular  our definition of the final mass  is the real mass of
the white  dwarf star  at the cooling  curve and  not the mass  of the
hydrogen free  core at the first  thermal pulse.  As it  was shown by
Kalirai  et al.   (2014) the  mass of  the core  increases  during the
TP-AGB, resulting  in a final  mass larger than  the one at  the first
thermal pulse. Therefore we  select theoretical computations that consider
the AGB evolution in the definition of the IFMR to compere with our results.

\begin{figure}
\begin{center}
\includegraphics[clip,width=0.50\textwidth]{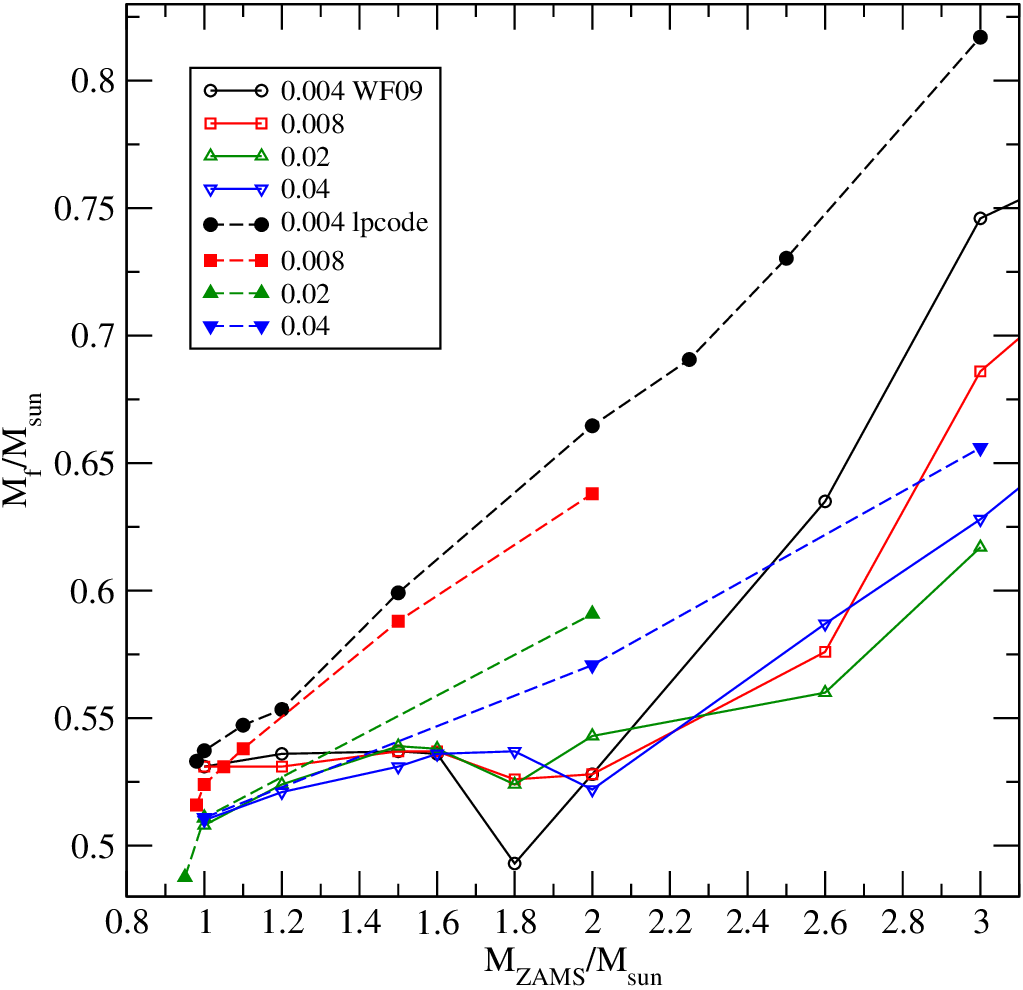}
\caption{Theoretical Initial-to-Final mass relation for sequences with
  metallicity  values   $Z=0.004$  (black  circles),   $Z=0.008$  (red
  squares), $Z=0.02$ (green triangles  up) and $Z=0.04$ blue triangles
  down). Hollow symbols correspond to the results of Weiss \& Ferguson
  (2009)  while fill symbols  correspond to  results obtained  in this
  work --  see text  for details. We  consider the initial  mass range
  where the two  studies overlap.  (A color version  of this figure is
  available in the on-line journal.) }
\label{Mi-Mf-WF09}
\end{center}
\end{figure}

\begin{figure}
\begin{center}
\includegraphics[clip,width=0.50\textwidth]{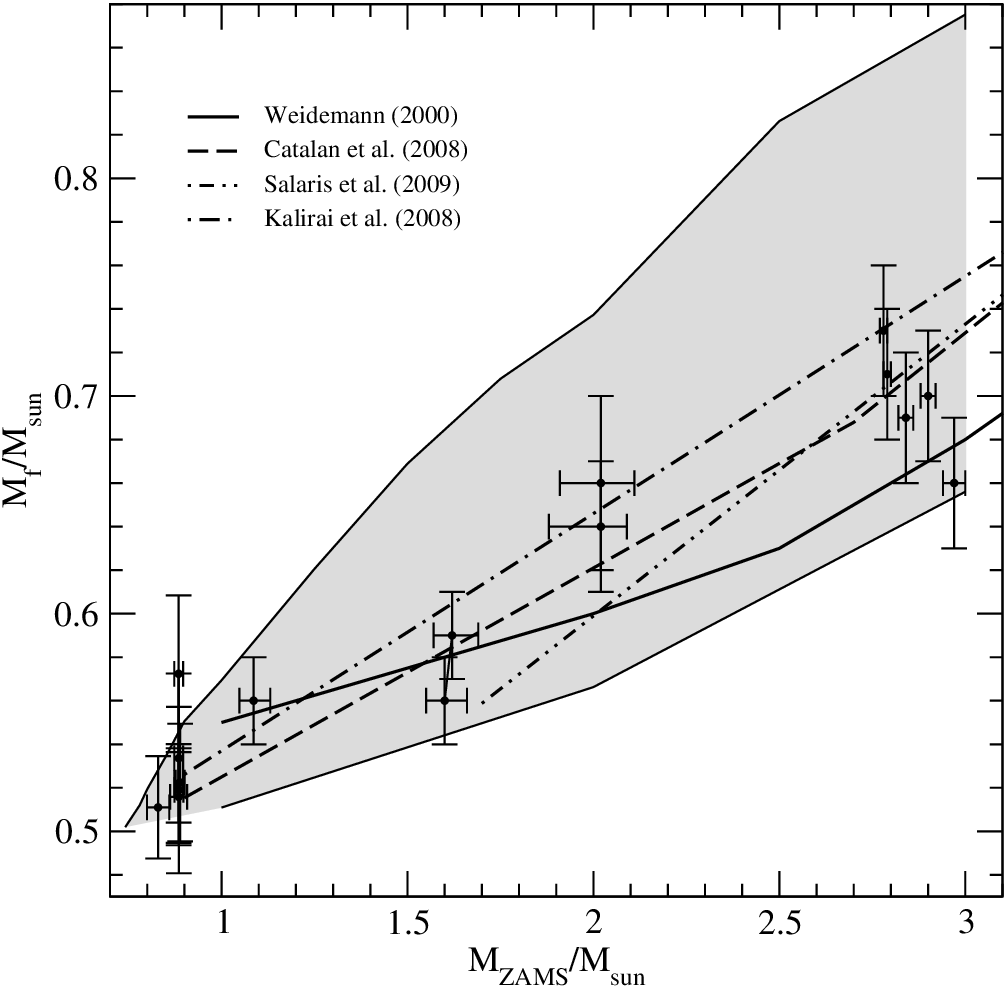}
\caption{Theoretical Initial-to-Final mass relation computed in this work (gray area) compare to the semi-empirical relations from Weidemann (2000, full line), Catal\'an et al. (2008, dashed line), Salaris et al. (2009, dashed-dot-dot line) and Kalirai et al. (2008, dashed-dot line). }
\label{empirica}
\end{center}
\end{figure}

Weiss \&  Ferguson (2009) computed sequences from  the ZAMS to the end of  the TP-AGB for 5 values  of metallicity between $Z=0.0005$ and $Z=0.04$  and initial masses  ranging from 1.0 $M_{\odot}$  to 6.0$M_{\odot}$.  We compare our results with those of Weiss \& Ferguson (2009)  in Fig. \ref{Mi-Mf-WF09}. Only  the metallicity values and  the initial  mass range  where both studies  overlap are  plotted.  At the  low  mass end  we  have  good agreement,  specially for  high metallicity  progenitors.  Differences are  more   important  for  high   mass  stars  and   low  metallicity progenitors. In their calculation, envelope overshooting was allowed during the TP-AGB, with $f=0.016$ for all sequences. This overshooting prescription leads to an enhanced reduction of the final core mass, leading to  an IFMR very close to the  relation between  the initial  mass and the  core mass  at the first thermal  pulse. Note the minimum  in the final mass  for Weiss \& Ferguson (2009)  around $\sim 2 M_{\odot}$ corresponding  to the limit between the sequences that experience an helium flash at the beginning of  the central  helium  burning phase,  and  those that  do not (see section \ref{dependence}).  This minimum  is also  present in  our computations  when we consider the  mass of  the hydrogen free core at the  first thermal pulse as the final mass (see Renedo et al. 2010).

Other authors (e.g. Karakas 2010; Marigo \& Girardi 2007; Marigo et al. 2013) have computed detailed atmosphere models of TP-AGB evolution, considering parametrized overshooting treatments and also a detailed composition of the surface. They calibrate the dredge--up onset and efficiency  with observations of carbon surface abundances of AGB stars. Because in Karakas (2010) and Marigo \& Girardi (2007) the authors consider dredge--up episodes during the TP-AGB we expect final mass for sequences with initial masses above $\sim 2M_{\odot}$ to be higher in our computations. We showed in section \ref{AGB} that for $Z\sim 0.02$  our final masses can be overestimated by less than 10\%, while for lower metallicity this difference can reach $\sim 20\%$. 

Karakas (2010) computed a large grid of TP-AGB models covering the metallicity range between $Z=0.0001$ and $Z=0.02$, and initial masses between 1 and $6M_{\odot}$. She considered a detailed treatment of the third dredge--up in order to reproduce the atmospheric carbon abundances observed in field stars. As we expected, for initial mass larger than $\sim 1.8-2.0M_{\odot}$ we obtain final masses larger than those of Karakas (2010), up to a $\sim 12\%$ for $Z=0.0001$ and $M_{\rm initial} = 3M_{\odot}$. For low initial masses our models show significantly lower final masses than that of Karakas (2010), up to $\sim  27\%$ for $1M_{\odot}$ and $Z=0.0001$. This large difference can be mainly due to the lack of mass loss during the pre-AGB evolution adopted for the initial models.  As in Wiess \& Ferguson (2009),  we consider mass loss during the RGB phase modeled by the Reimers formulae (Reimers 1975). Therefore our sequences begin the AGB evolution with a smaller mass than that of the ZAMS, limitng also the number of thermal pulses and thus, the growth of the hydrogen free core during the TP-AGB.
Similar results to those of Karakas (2010) are found when we compare our IFMRs with those of Marigo  \&  Girardi  (2007). They  also  computed detailed  models  for  TP-AGB evolution for $Z=0.0001$ to $Z=0.03$ and $M_{\rm initial} = 0.5-5.0 M_{\odot}$, calibrating the onset and the efficiency of the third dredge--up with observations, such as carbon star luminosity functions in the Magellanic Clouds and TP-AGB lifetimes in Magellanic star clusters. The initial models at the first thermal pulse are taken from Girardi et al. (2000), derived again from constant mass computations. They still find a significant dependence with metallicity, leading to an allowed final mass range of $0.1 M_{\odot}$ wide for a given initial mass (see their figure 26).

Finally, Kalirai et al. (2014) found a grid of best fit models from the model grid of Marigo et al. (2013), therefore without mass loss in the RGB, for $Z=0.02$ that better represents the core mass growth during the TP-AGB for solar metallicity in the initial mass range of $1-4.4M_{\odot}$. The adopted dredge--up efficiencies varies with the initial mass. From their best-fitting TP-AGB models we can extract an IFMR (see their Table 3) and compare their predictions with ours. As a result, we found that the final masses from Kalirai et al. (2014) are  somewhat larger than ours in the whole initial mass range. The differences are small, up to $\sim 4\%$ compared to our $Z=0.01$ grid, and $\sim 8\%$  as compared to our $Z=0.02$ models. The smaller differences are for initial mass around $2.5 M_{\odot}$, where dredge--up efficiencies are larger for this metallicity.

In  addition  to  theoretical  computations, several  studies  where
dedicated to determine a semi-empirical IFMR from observations of star
belonging  to globular  and open  clusters and  binary systems  on the
Galactic  field. We selected  the semi-empirical
IFMR  from Catal\'an  et  al.   (2008b), Salaris  et  al.  (2009)  and
 Kalirai et al. (2008) to compare with our theoretical results. We also
include the classical semi-empirical  IFMR from Weidemann (2000). The
results are  shown  in   Figure  \ref{empirica}.  Most of the
semi-empirical results fall between our theoretical IFMR characterized
by  $Z=0.004$  and  $Z=0.02$.  At   the  low  mass  end,  below  $\sim
1.7M_{\odot}$, the IFMR of Catal\'an et al. (2008b) and Kalirai et al. (2009)
fall between the $Z=0.004$ and $Z=0.01$ curves. Salaris et
al. (2009) IFMR  is better matched by our $Z=0.02$ curve. At the high
mass end, the semi-empirical IFMR are better
matched  by  considering  metallicity  values  between  $Z=0.008$  and
$Z=0.01$. This  is a consequence of  the sample used  to determine the
IFMR, corresponding  to younger, and more metallic  open clusters, and
common  proper motion pairs  from the  solar vicinity.  In particular,
Kalirai et al. (2008) extended  the low mass end towards lower initial
masses by including  the old metal-rich open cluster  NGC 6791. In all
cases,  the  semi-empirical  IFMR  demonstrate that  lower  mass  main
sequence stars produce lower mass white dwarf stars (Kalirai 2013).

\section{White dwarf models and the metallicity dependence}
\label{wd-parameters}

\subsection{Hydrogen envelope thickness}
\label{H-env}

\begin{figure}
\begin{center}
\includegraphics[clip,width=0.48\textwidth]{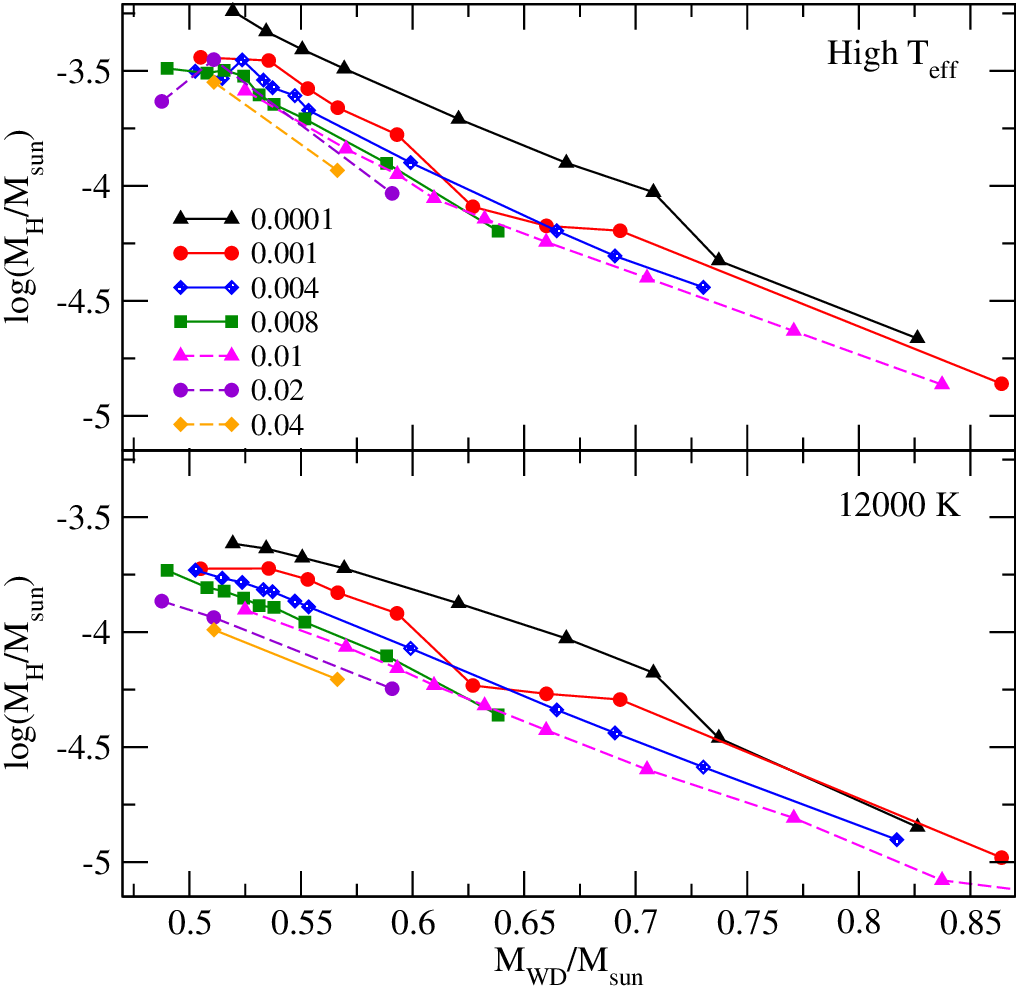}
\caption{Hydrogen mass at the cooling sequence as a 
function of the white dwarf mass, at the point of maximum effective temperature at the beginning of the white dwarf cooling branch (top panel) and for models at late stages on the white dwarf cooling curve, with effective temperature $\sim 12\, 000$ K. Each curve correspond to a different initial metallicity (see references on the plot). A color version of this figure is available in the on-line journal.}
\label{env}
\end{center}
\end{figure}

As shown by Iben \& MacDonald (1986) --see also Renedo et al.  (2010);
Miller  Bertolami et al.  (2013)-- the  amount of  hydrogen left  in a
white dwarf  star is metallicity  dependent, being thicker  for models
with less  metallic progenitors.  As a result  of the  larger hydrogen
envelopes, residual H  burning is expected to become  more relevant in
white  dwarf stars  with  low metallicity  progenitors.  Results  from
Miller  Bertolami (2013)  showed  that for  metal  poor stars  ($Z\sim
0.0001$), the  luminosity is completely dominated  by nuclear burning,
even at rather low luminosities.  This leads to a significant delay in
the  cooling times, as  compared to  stars with  higher metallicities,
such as  $Z=0.01$, in  which nuclear burning  does not play  a leading
role,  and most of  the energy  release come  from the  thermal energy
stored in the interior.  From the top panel of Figure \ref{env}, where
we show  the total  hydrogen content  in solar units  in terms  of the
white dwarf mass at the point of maximum effective temperature, we see
that,  in general,  a lower  hydrogen content  is found  for  a higher
metallicity  value for the  progenitor star.   Some departures  of the
expected trend can be noticed from this figure also. For instance, the
change in  the slope  for high  mass $Z=0.0001$ models  is due  to the
occurrence of the third dredge--up  for the two more massive models, as
also shown in Figure 1 of Miller Bertolami (2013).  For $Z=0.001$, the
hydrogen content for the 0.627$M_{\odot}$  model is lower than that of
the Z=0.004.  The largest departures  from the general trend are found
for low  mass white  dwarfs, in particular  those with  stellar masses
below $\sim  0.53 M_{\odot}$. As we mention  in sec. \ref{model-grid},
sequences  with the  lowest  masses of  the  grid experience  hydrogen
sub-flashes before  entering its  final cooling track.  In particular,
the sequence  with initial metallicity  $Z=0.02$ and white  dwarf mass
$0.488 M_{\odot}$ experience a  late thermal pulse which significantly
reduce the amount of hydrogen left in the star (Althaus et al. 2005b).

As  residual hydrogen  burning is  active,  the mass  of the  hydrogen
content decreases,  until it reaches  a certain value ($\sim  8 \times
10^{-5} M_{\odot}$  for solar metallicity  stars, Renedo et  al. 2010)
where the pressure  at the bottom of the envelope  is not large enough
to support further nuclear reactions, and the hydrogen content reaches
a  stationary value.  In addition,  time-dependent  chemical diffusion
builds a pure hydrogen envelope  when the star reaches lower effective
temperatures ($\sim 30\, 000$ K  for solar metallicity). Then, at late
stages of the cooling, a smoother distribution of the hydrogen mass in
terms of the stellar mass and metallicity can be expected. This can be
seen for  our model  grid from the  bottom panel of  Figure \ref{env},
where we  depict the total  hydrogen mass as  a function of  the white
dwarf mass at  $\sim 12\, 000$ K. In particular, at  the low mass end,
the  hydrogen   envelope  mass   clearly  decreases  as   the  initial
metallicity increases.

A comparison between the resulting hydrogen content for a fixed white dwarf mass
shows a spread with metallicity. For instance, for a white dwarf with a  stellar mass of $\sim
0.570 M_{\odot}$ we found that the reduction of  the hydrogen content
due  to an  increase  in  the initial  metallicity  from $Z=0.001$  to
$Z=0.04$ can be up to a factor  of 2.  This difference is small and do
not introduce a measurable difference on the cooling time scale at low
luminosities.   However we must  take into  account that  the hydrogen
content resulting from canonical evolutionary computations is a higher
limit.  Asteroseismological  studies show that the  mean hydrogen mass
for pulsating ZZ  Cetis stars is $\sim  10^{-6}M_*$, around two
orders of  magnitude lower than the  canonical values for  most of the
sequences computed in this work (Castanheira \& Kepler 2009; Romero et
al. 2012, 2013).

\subsection{Central oxygen abundances}

\begin{figure}
\begin{center}
\includegraphics[clip,width=0.48\textwidth]{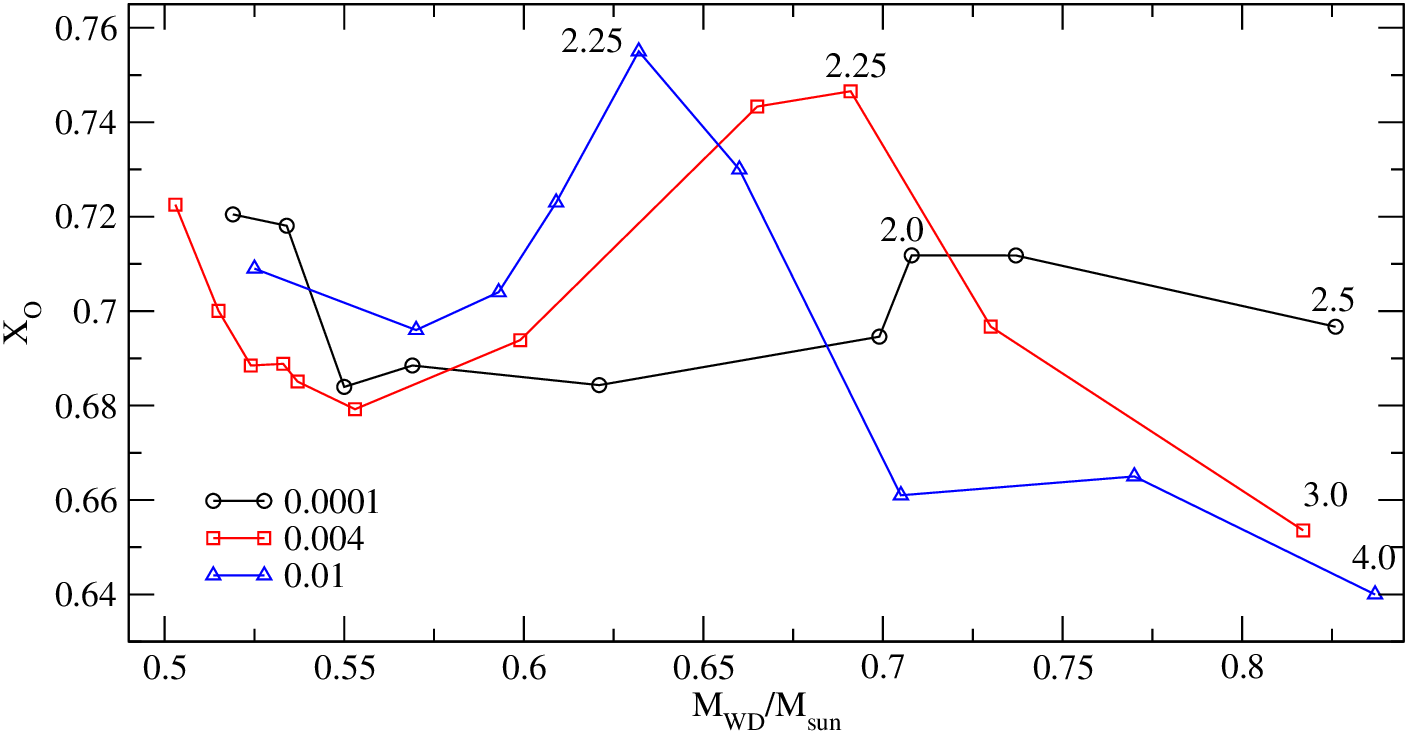}
\caption{Central oxygen abundance in terms of the white dwarf mass, for three metallicities, $Z=0.0001$ (black circles), $Z=0.004$ (red squares) and $Z=0.01$ (blue triangles). The values are taken after rehomogeneization but before crystallization sets in the central core. The numbers indicate the progenitor mass at the ZAMS for some points.}
\label{central-o}
\end{center}
\end{figure}

Figure \ref{central-o}  displays the central oxygen  abundance by mass
 left  after core helium burning  in term of  the white dwarf
mass   for  three   metallicity  values   $Z=0.0001$,   $Z=0.004$  and
$Z=0.01$. The  numbers in the figure indicate  the mass of  the progenitor  star for
some points. The  predictions of our calculations for  $Z=0.01$ are in
good  qualitative  agreement  with  the  computations  of  Salaris  et
al.   (1997)  for   solar  metallicity,   as  shown   in   Althaus  et
al. (2010). We recall that the final carbon$-$oxygen stratification of
the  emerging  white dwarf  depends  on  both  the efficiency  of  the
$^{12}$C$(\alpha,  \gamma )^{16}$O  reaction rate  and the  occurrence of
extra$-$mixing  episodes   toward  the   late  stages  of   core  helium
burning.  Note that  for all  sets, a maximum  in  the oxygen
abundance is predicted, being  the mass of the  white dwarf model  larger for lower
metallicity  values. However,  when we  consider the  initial  mass, we
found  that the  maximum is  located approximately  at the  same value,
around   $2.25M_{\odot}$  for  $Z=0.004$   and  $Z=0.01$   and  around
$2M_{\odot}$ for  $Z=0.0001$. This maximum is also  present in Salaris
et al.  (1997) models. After  that, the oxygen abundance  decreases and
tends to a  stable value oscillating between 0.61 and  0.63, as also shown in
Romero et al.  (2013) for white dwarf models  with $Z=0.01$. 
Finally, note that the changes in the oxygen and carbon central 
abundances due to metallicity can get up to $\sim 10\%$ in the metallicity range 
shown in Fig. \ref{central-o}.

\section{The age of stellar populations and the metallicity parameter}
\label{age}
        
As it is known, the age of the model before
entering the  final cooling sequence  is metallicity dependent; for a
given progenitor mass  on the ZAMS, the age of the  pre-WD star at the
point  of   maximum  effective   temperature  is  significantly smaller   for  lower
metallicities.  For  progenitor stars corresponding to initial masses  
above $\sim  1.5 M_{\odot}$,  the evolutionary
scales are small, from 2 Gyr to  hundreds of Myr. But for low mass
stars, those that present a fraction of its core in a degenerate state
after central  hydrogen exhaustion,  the evolutionary scales  are much
longer, and  can reach 15  Gyr. In addition,  for low mass  stars the
time spent  on the  RGB phase is  not negligible  and can reach  up to
$\sim 2-3$ Gyr for $\sim 1M_{\odot}$. Thus the metallicity dependence of the pre-WD age will
be more  important for low  mass sequences and older populations
like globular  clusters. We can see from  Table \ref{tabla-masa} that the
pre-WD age  of a  1$M_{\odot}$ progenitor  star is reduce  from 12.5
Gyr  to  5.9 Gyr  when  the metallicity  goes  from  a solar  value
($Z=0.02$) to very metal poor values characteristic of halo population
($Z=0.0001$).

\begin{figure}
\begin{center}
\includegraphics[clip,width=0.50\textwidth]{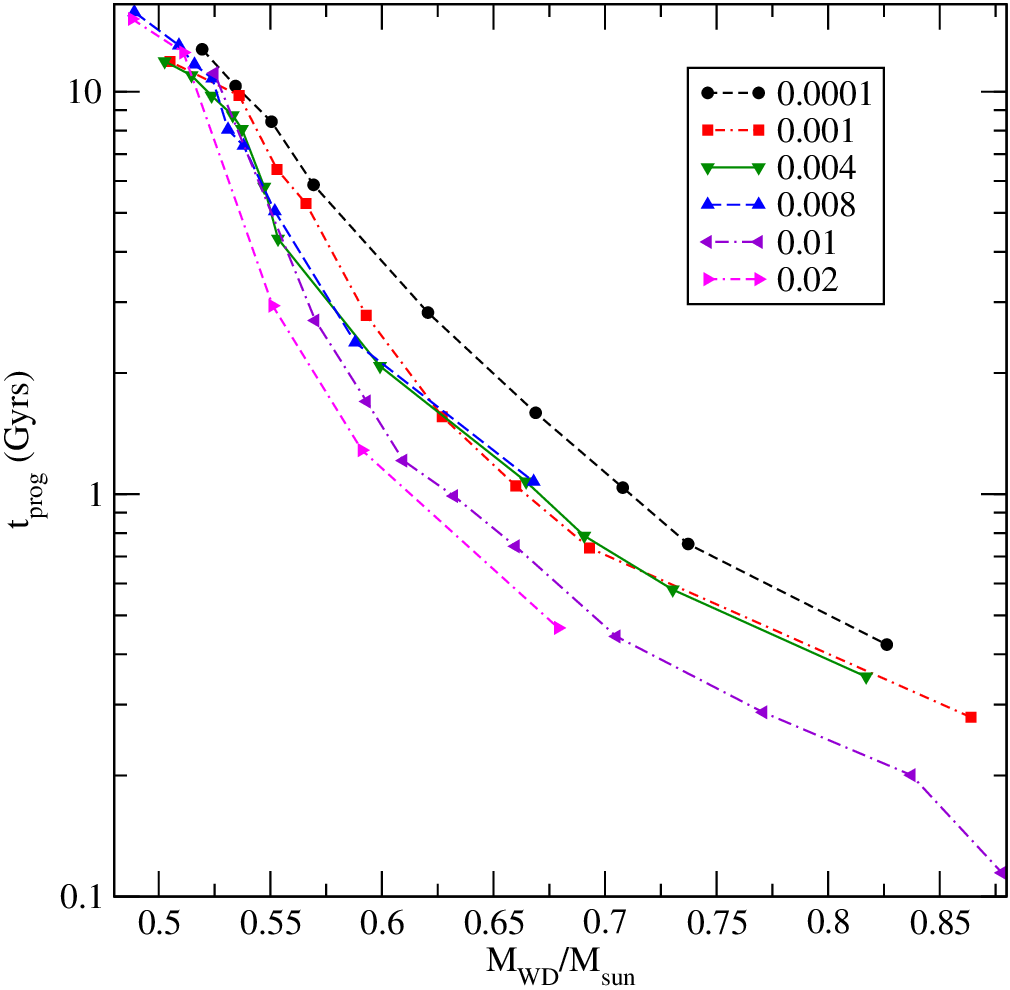}
\caption{Lifetime previous to the white dwarf stage versus stellar mass on the cooling curve. Each curve corresponds to a single initial metallicity. (A color version of the figure is available on the on-line version).  }
\label{Mf-time}
\end{center}
\end{figure}

Figure \ref{Mf-time} shows the pre-WD age ($t_{\rm prog}$) in terms of the
white  dwarf  mass  for  all  metallicity values  considered  in  this
work. The general trend shows that  the age, for a given stellar mass,
decreases  with  metallicity (e.g. Isern et al. 2005). When the
step on metallicity is sufficiently small, we can notice a departure from the general trend, caused by the difference in the progenitor mass.  As it
was shown in section \ref{i-f}, the  mass of the resulting white dwarf star
decreases      when     metallicity      increases      (see     Table
\ref{tabla-masa}). This is a consequence of an enhanced mass loss rate
with increasing  metallicity during the  giant stages.  Note  that the
metallicity dependence of  the final mass will still  be present if we
were to consider the mass of the hydrogen  free core at the
first thermal  pulse, since the  mass of the carbon-oxygen  core after
helium  burning increases  with  decreasing metallicity,  as shown  in
Figure \ref{Mi-Mf-1}.   Two effects are  present having opposite
influences.  For  a  fixed  white  dwarf mass,  the  lifetime  of  the
progenitor star decreases with metallicity but, on the other, the mass
of the progenitor is smaller  and its lifetime increases (e.g. Isern et al. 2005). For a low mass
white dwarf, the age difference can be as high as 8 Gyr.

\begin{figure}
\begin{center}
\includegraphics[clip,width=0.50\textwidth]{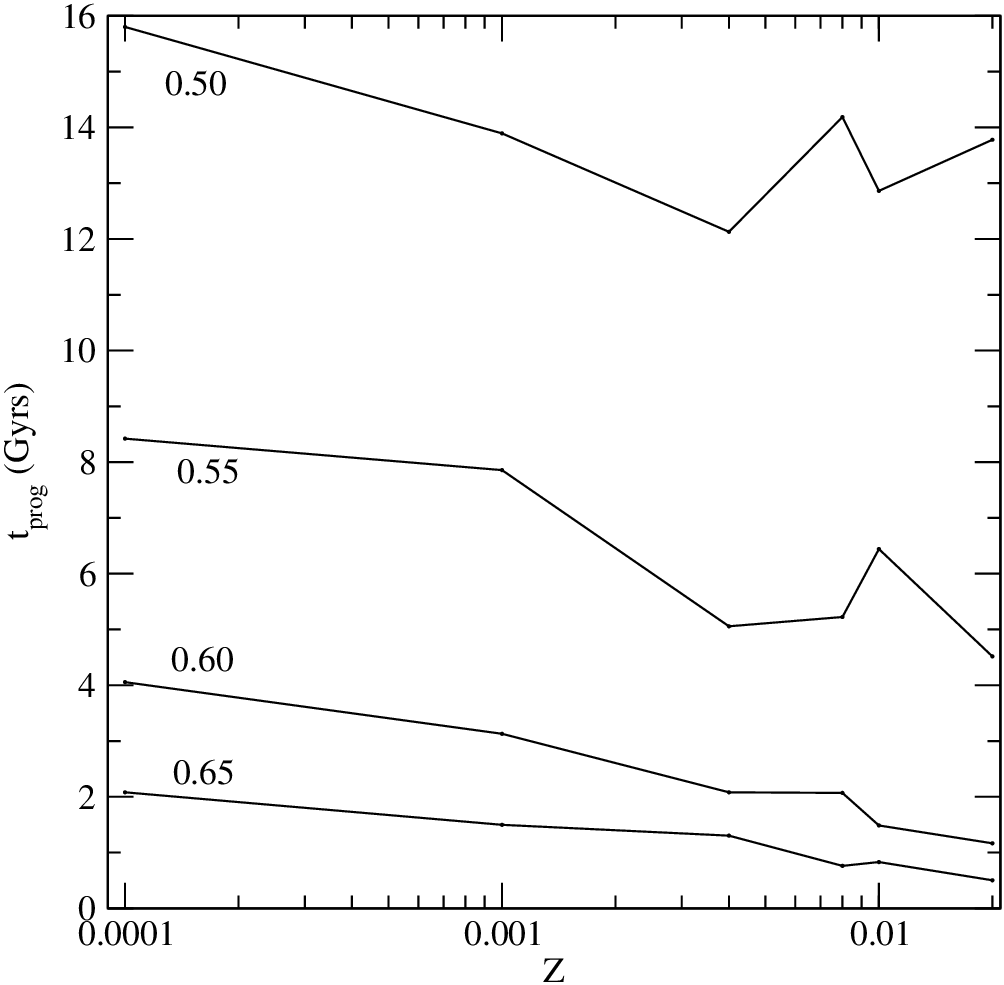}
\caption{Pre-WD lifetime in terms of the initial metallicity for white dwarf models with masses 0.50, 0.55, 0.60 and 0.65$M_{\odot}$. }\label{time-Z}
\end{center}
\end{figure}

In Figure \ref{time-Z} we show the age of  the progenitor star in terms of the
initial  metallicity. From this figure the differences on the  white dwarf  total  age due  to metallicity  is
clearer. We found that  the age decrease for low $Z$  but the
dependence  is  not  always  monotonous,
indicating the influence of  another  parameter,   i.e.   the  IFMR.  
As the  stellar  mass increases,  the  pre-WD lifetime is
shorter, and  usually negligible, as  compared to the cooling  time on
the white dwarf sequences once they evolved to lower luminosities.

We can do  a simple exercise by estimating the total  age of a typical
halo   globular    cluster.   Kalirai   et    al.   (2009)   determined
spectroscopically that  the mass for  the bright white dwarf  stars in
M4, at the  beginning of the observed cooling  sequence, is $0.592 \pm
0.012 M_{\odot}$. Assuming that we  do not know the metallicity of the
cluster, we can  estimate the age of the progenitor,  and of the star,
from the pre-WD ages listed in Table \ref{tabla-masa}. The results are
listed  in  Table \ref{ex},  along  with  the  estimated mass  of  the
progenitor star.  Note that all  progenitor masses are low  mass stars
and do not experience many thermal pulses, between 1 and 5, during the
AGB. The  pre-WD age  does not decrease  monotonously, as we  saw from
Figs.  \ref{Mf-time}  and  \ref{time-Z}.  For  the  metallicity  value
estimated spectroscopically for M4,  $Z\sim 0.001$, the estimated values
of progenitor  mass for the bright  white dwarfs is  in agreement with
the estimations  of the  stellar mass of  the main sequence  turn off,
$\sim 0.85 M_{\odot}$ (see Sec. \ref{cluster}).

\begin{table}
\centering
\caption{Age and progenitor mass for different metallicities for a white dwarf star with mass $0.592 \pm 0.012 M_{\odot}$, estimated by Kalirai et al. (2009) for the bright white dwarf stars of M4. }
\begin{tabular}{ccc}
\hline\hline 
 $Z$ & Age [Gyr] & $M_{ZAMS}/M_{\odot}$ \\
\hline
0.0001 & 11.133 & 0.83\\
{\bf 0.001}  & {\bf 10.587} & {\bf 0.90}\\
0.004  & 9.196  & 0.97\\
0.008  & 8.834  & 1.03\\
0.01   & 10.369 & 1.04\\
0.02   & 7.499  & 1.36\\
0.04   & 9.413  & 1.22\\  
\hline
\label{ex}
\end{tabular}\\
\end{table}

\subsection{The white dwarf cooling curve in globular clusters}
\label{cluster}

\begin{figure*}
\centering                  
\includegraphics[clip,width=0.45\textwidth,angle=0]{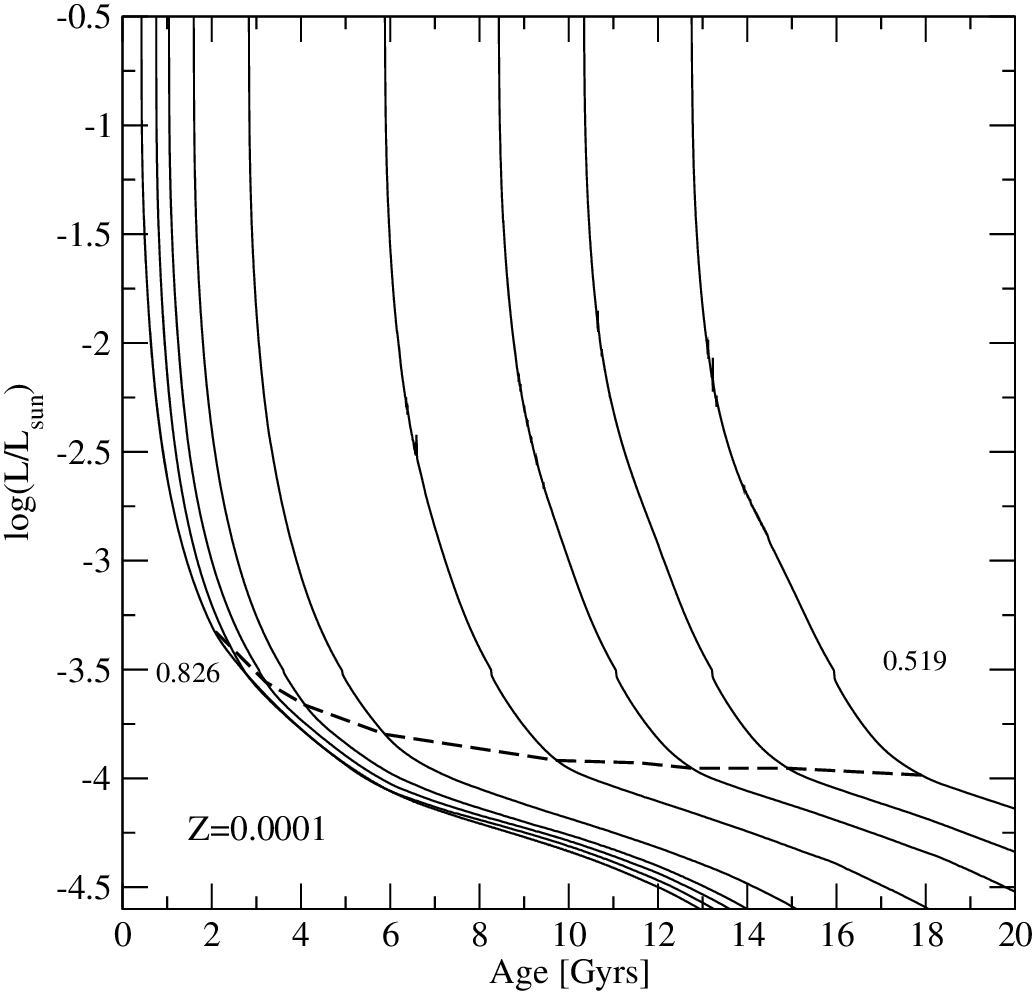}    
\includegraphics[clip,width=0.45\textwidth,angle=0]{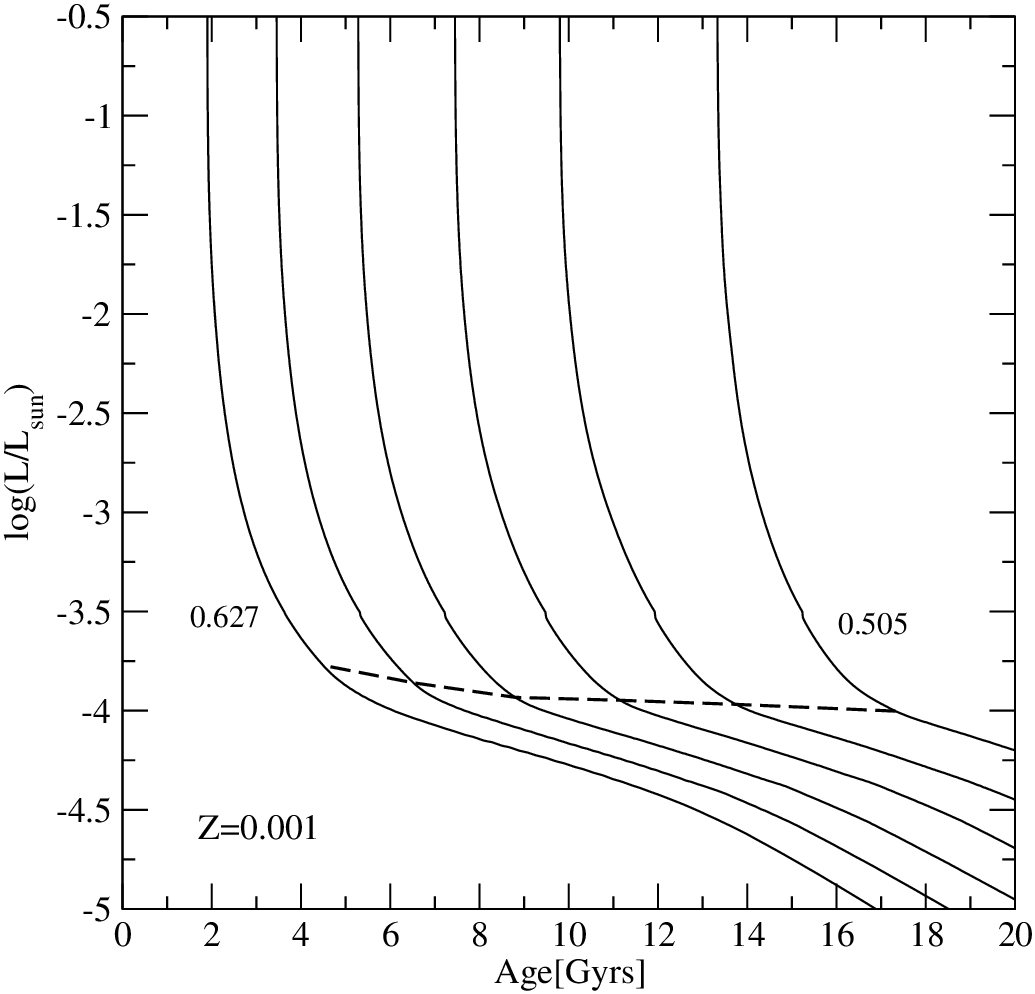}
\includegraphics[clip,width=0.45\textwidth,angle=0]{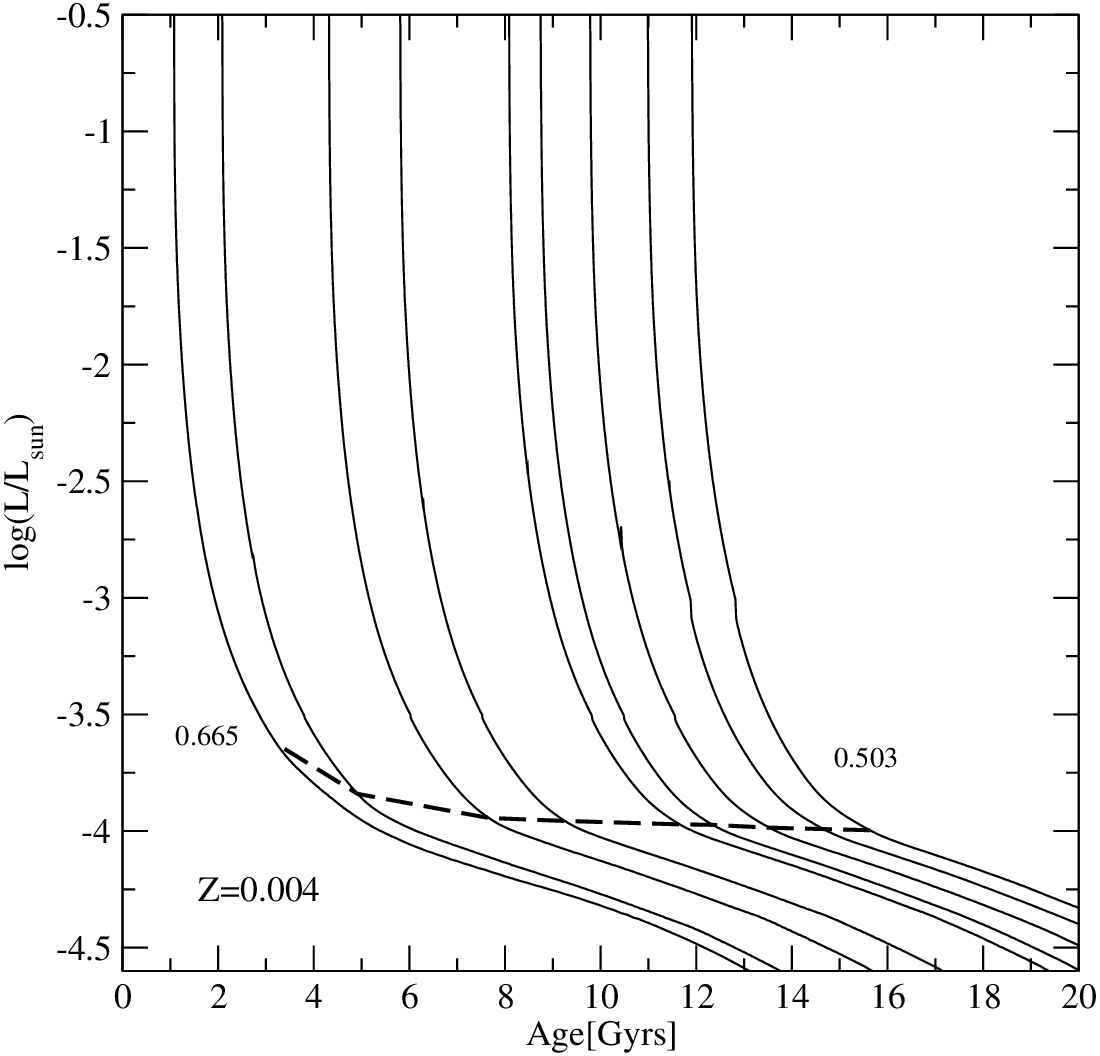}
\includegraphics[clip,width=0.45\textwidth,angle=0]{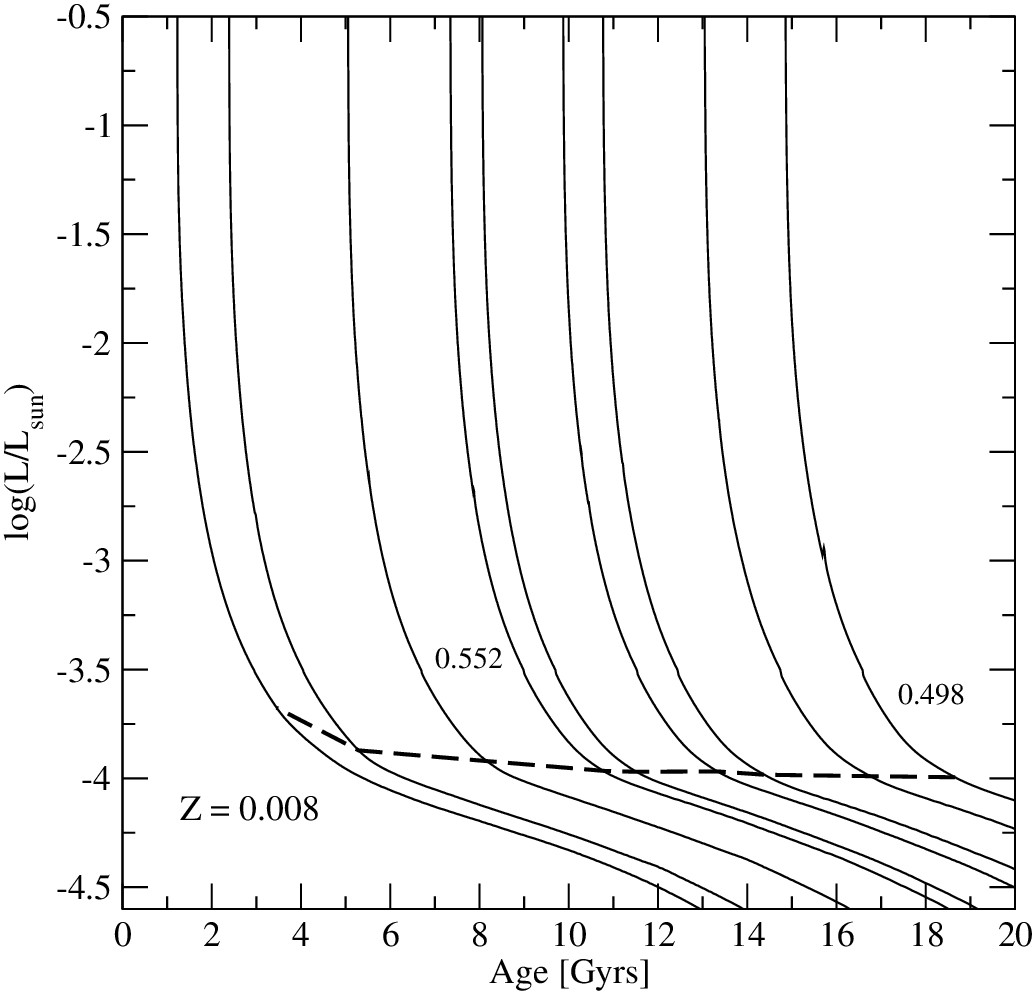}
\caption{Total  luminosity  as  a  function  of age  for  our  set  of
  sequences with  four different  metallicity of the  progenitor star.
  The age  corresponds to the total  age of the  model, by considering
  all the stages  previous to the white dwarf  cooling. The horizontal
  dashed  line  represents  the  point  in  the  evolution  where  the
  crystallization process begins at the core, by means of the Horowitz
  et al.  (2010) phase  diagram for crystallization.  The stellar mass
  values in solar  mass units, from left to  right, are: o.519, 0.534,
  0.550,  0.561,  0.569, 0.621,  0.669,  0.708,  0.737  and 0.826  for
  $Z=0.0001$, 0.505, 0.536, 0.553,  0.566, 0.593, 0.627 for $Z=0.001$;
  0.503, 0.515, 0.524, 0.533, 0.537, 0.547, 0.553, 0.599 for $Z=0.004$
  and  0.498, 0.509,  0.516,  0.524, 0.531,  0.538,  0.552, 0.588  for
  $Z=0.008$.} 
\label{cooling}
\end{figure*}

In Figure \ref{cooling} we show  the total luminosity as a function of
age  during  the white  dwarf  cooling  sequences  for four  different
metallicity  values $Z=0.0001,  0.001, 0.004$  and $0.008$.   For each
sequence we consider the total age  of the star, i.e. the age measured
from  the ZAMS  considering all  previous  stages to  the white  dwarf
cooling curve, added to the cooling lifetime.  The mass range for each
metallicity is indicated in  the plot.  Finally, the horizontal dashed
line represents  the point in the cooling  curve where crystallization
in the  central core begins, following of  the Horowitz et  al.  (2010)
prescription  of  the  C$-$O phase   diagram.   The  white  dwarf  sequences
characterized by  a progenitor metallicity of  $Z=0.001$ and $Z=0.01$,
originally  computed in Renedo  et al.   (2010), were  recalculated by
considering the Horowitz et al.  (2010) phase diagram.  Note that with
the  Segretain \& Chabrier  (1993) phase  diagram in  the evolutionary
computations, the effective temperature at which crystallization begins
at the center of the star  is $\sim$ 1000 K higher than that predicted
by  Horowitz et al.   (2010) (see  Romero et  al.  2013  for details).
Although  white  dwarf  atmospheres  at  late stages  of  the  cooling
sequences are metal free, the final white dwarf evolution is not 
insensitive to the metallicity of the progenitor star. As we showed in
previous  sections,  the pre-WD  lifetime  is  larger  for metal  rich
sequences as  compared to low  metallicity progenitors.  In case  of a
$1M_{\odot}$ progenitor  star, the  progenitor lifetime goes  from 5.9
Gyr  to  11.1  Gyr  when  the  initial  metallicity  increases  from
$Z=0.0001$  to  $Z=0.01$.   On  the   other  hand,  as  we  showed  in
sec. \ref{H-env}, the total amount of hydrogen left in the white dwarf
remnant  is  usually larger  for  lower  metallicities,  so a  thicker
hydrogen layer is expected for low $Z$ sequences.

Our  lifetime  computations  during  the  white  dwarf  sequences  are
consistent  to  those  of  BaSTI   evolutionary  code  up  to  2\%  at
luminosities  lower than  $\log (L/L_{\odot})  \sim -1.5$  (Salaris et
al.  2013).  This difference  is  smaller  than  the uncertainties  in
cooling times  attributable to the present uncertainties  in the white
dwarf  chemical stratification.  We  also computed  sequences with  an
initial mass $1 M_{\odot}$  and metallicity of $Z=0.0001$ and $Z=0.02$
using MESA (Paxton et al.  2011, 2013) and found lifetimes consistent
with ours, with differences of $\sim 0.1$ Myr at $\sim 25\, 000$ K on the cooling curve.

HST observations of  globular clusters  in the
Galactic Halo, M4 (Richer et al.   2004; Hansen et al. 2004), NGC 6397
(Richer  et al.   2006; Hansen  et al.   2007) and  47 Tuc  (Hansen at
al. 2013),  have uncovered several  hundred member white  dwarf stars.
From a sample of galactic  globular clusters, Krauss \& Chaboyer (2003)
derived  an  mean age  of  $12.6^{+3.4}_{-2.2}$  Gyr. More  recently,
Kalirai (2012) derived an age of the stellar halo near the position of
the Sun  to be  $11.4\pm 0.7$ Gyr.   Consequently, given  these large
ages, bright white dwarf  stars are expected to have  low stellar mass values
around $0.51-0.55  M_{\odot}$ (Renzini \& Fusi Pecci  1988; Renzini et
al. 1996).  With  the high precision data available  we can employ the
current observational constrains on  age to derive some properties for
the white  dwarf sample in these  systems.   In particular, we
used  the  total age  determinations  of  three  globular clusters  to
estimate  the mass  range on  the white  dwarf cooling  curve  and the
minimum  mass  expected  to  have  some  fraction of  its  core  in  a
crystallized state. 

The globular cluster NGC 6397  is one of the closest globular clusters
(2.6 kpc)  and is  also a metal  poor system  with a iron  to hydrogen
ratio of [Fe/H]$=-2.00\pm 0.01$, corresponding to $Z=0.0001$. For this
globular  cluster Hansen  et  al.  (2013) determined  a  total age  of
11.7$\pm$0.3 Gyr.  Considering this total age we found that the white
dwarf stars at high luminosities at the beginning of the cooling curve
are  characterized  by  a  mean  mass of  $0.535\pm  0.008  M_{\odot}$
corresponding to  an initial  mass of $\sim  0.85 M_{\odot}$.   On the
other hand, the  corresponding minimum stellar mass for  a white dwarf
to have undergone  some degree of crystallization in  the central core, 
as seen by Winget et al. (2009, 2010), is  $0.561\pm  0.007  M_{\odot}$,  
with  a progenitor  mass  of  $\sim
0.95M_{\odot}$. The  observed white dwarf  mass is not the  same along
the cooling  curve. White  dwarf stars with  higher stellar  mass cool
faster, having lower  luminosity values at a given  age.  In addition,
the   pre-white  dwarf  lifetime   decreases  dramatically   when  the
progenitor mass increases, so these  stars reach the white dwarf stage
in a shorter amount of time.

For M4, Hansen et al.  (2004) derived an age of 12.1 Gyr, with a 95\%
lower bound  of 10.3  Gyr, by direct  fitting of the  Color--Magnitude
Diagram.   Bedin et  al.  (2009)  also  determine the  cluster age  of
$11.6\pm  0.6$ Gyr, consistent  with the  age from  fits to  the 
main--sequence turnoff  (12.4$\pm$1.4 Gyr).  The iron  to hydrogen fraction
for   M4  is   [Fe/H]$=-1.1\pm  0.01$   (Mucciarelli  et   al.   2011)
corresponding  to  $Z=0.0013$.  We  consider  that  the grid  of
sequences  with  Z=0.001  is   representative  of  this  cluster.   By
considering the age determinations from  Bedin et al.  (2009) we found
that the  mass at the top  of cooling curve is  around 
$\lesssim 0.536 \pm  0.015 M_{\odot}$, in  agreement with  the mass  determinations of
Kalirai et al.  (2009) of $0.529 \pm 0.012  M_{\odot}$.  Note that the
progenitor mass for a white dwarf with $\sim 0.536 M_{\odot}$ is 
$\sim 0.92 M_{\odot}$  at $Z=0.001$.   Also, those sequences  with white dwarf
mass $\gtrsim 0.553 \pm  0.008 M_{\odot}$ have reached crystallization
temperatures at  an age $\sim  11.6$ Gyr.  So sequences  with initial
mass  larger than  1$M_{\odot}$  are already  crystallized. The  small
difference between the age derived by  Bedin et al. (2009) and that of
Hansen et al. (2004) do not change considerably our results.  

The globular cluster 47 Tuc is  more metal rich than both NGC 6397 and
M4,  with [Fe/H]=-0.75  (Carretta  et  al.  2009),  so  we use  a
metallicity value of $Z=0.004$ for this cluster.  Hansen et al. (2013)
determine a cluster age by fitting the properties of the cluster white
dwarf population.   They derived  a total age  of 9.9$\pm$0.7  Gyr at
95\%  confidence.  From our  computations we  found that  the sequence
characterized  by  a white  dwarf  mass  of  $0.537M_{\odot}$ has  not
reached  the  crystallization  temperature (see  Fig.   \ref{cooling})
while  all  sequences with white dwarf mass  $\geq 0.547M_{\odot}$,  and
progenitor mass of $\geq  1.1M_{\odot}$, have a partially crystallized
core at 10  Gyr. Finally the white dwarf stars  at high luminosity in
the cooling curve  are characterized by an stellar  mass around $0.524
\pm  0.005   M_{\odot}$  corresponding  to progenitors  with
$0.95M_{\odot}$.

Summarizing, the stellar mass values  of the white dwarf stars at high
luminosities derived for the three globular clusters are very similar,
around $\sim  0.53 M_{\odot}$,  in agreement with  previous estimates.
Nevertheless, the progenitor  mass at the ZAMS changes with metallicity.   
The progenitor mass  for NGC
6397 is  around $0.85M_{\odot}$, while for M4  is $\sim 0.92M_{\odot}$.
Since the  total ages  for these two  systems are basically  the same,
within  the   uncertainties,  the  initial  mass   differences  are  a
consequence of  the different characteristic metallicities.  47 Tuc is
younger than NGC 6397 and M4 by  $\sim 1-2$ Gyr and thus the mass of
the main sequence is  larger ($0.95 M_{\odot}$), as expected.  However
the  stellar mass  at high  luminosities  in the  white dwarf  cooling
sequences is  the lowest  of the sample,  indicating an  enhanced mass
loss rate, associated also to the higher metallicity.
 
\section{Conclusions}
\label{conclusions}

In this work  we studied the impact of metallicity  on the theoretical
white dwarf  models, considering  also its  influence on  the previous
stages  of  evolution  through  the  analysis  of  the  IFMR  and  the
evolutionary  lifetimes.   To this  end  we  compute  a grid  of  full
evolutionary  sequences characterized  by metallicity  values for  the
progenitor stars between  $Z=0.0001$ and 0.04, covering  the $Z$ range
observed  on the  galactic disk  and halo  populations. We  focused on
stars  with   initial  masses  ranging  from   $0.85M_{\odot}$  to  $3
M_{\odot}$, giving  rise to  white dwarf stellar  masses in  the range
between $\sim 0.5M_{\odot}$ $\sim  0.87M_{\odot}$.  Our  main results  are the
following:

\begin{itemize}

\item From  our theoretical  evolutionary computations, we  found that
  the  IFMR  shows a  dependence  with  the  initial metallicity:  the
  stellar  mass  of the  remnant  on  the  white dwarf  colling  curve
  decreases as the initial metallicity increases, indicating an strong
  influence of the  metallicity on the mass loss  rates during the RGB
  and  the  AGB  phases.  Our  results  are  in  good  agreement  with
  observations and  semi empirical estimations  of the IFMR,  showing
  that the spread  observed in the semi empirical  data can be explain
  by means of the different metal abundances.

Our analysis on the IFMR does  not include the effect of extra--mixing episodes during the TP-AGB evolution. 
Then for most sequences dredge--up episodes, and its effects on the effective temperature, mass$-$loss rate, and core mass growth during
the  TP-AGB evolution, are not present.  It follows  that the  dependence of  the IFMR  is explored  mainly under  the
conditions of fixed initial  metallicity,  and  photospheric C/O ratio below unity during the TP-AGB phase, and thus no C--rich stars are formed from our models, except for sequences with high initial mass and metallicity $Z=0.0001$.

\item The  mass of  the helium  core at the  beginning of  the central
  helium  burning is  smaller for  more metal  rich models.  This will
  strongly  influence   the  determination  of   the  inner  chemical
  composition   of  white   dwarf  stars   with  stellar   mass  
$\sim  0.5M_{\odot}$. Finally, the minimum mass  for a star to undergo a
  helium flash decreases with metallicity.

\item We confirm the dependence  with metallicity of the hydrogen mass
  left at a white dwarf star, being in general thinner for models with
  higher $Z$ progenitors.

\item We found  that as the metallicity increases  the pre-white dwarf
  lifetimes  also  increases,  with age differences up  to  $\sim 5-6$  
  Gyr  for  a  $\sim 1M_{\odot}$ star.  Then,  in order to estimate the  age of a stellar
  population or a single object properly, we not only need to consider
  the  total pre-white  dwarf  lifetime  but also  we  must take  into
  account its characteristic metal content.
  
\item Finally we employed our evolutionary computations to study three
  old globular clusters  characterized with very different metallicity
  values:   NGC  6397   ($Z=0.0001$),  M4   ($Z=0.001$)  and   47  Tuc
  ($Z=0.004$). By  taking the age determinations  from the literature,
  we estimate the mass of the white dwarf at high luminosities and the
  corresponding mass  of the progenitor star, employing  the IFMR with
  the corresponding $Z$ value. We found that all three clusters have a similar
  white  dwarf mass  at high  luminosities ($\sim  0.53M_{\odot}$), in
  agreement with previous estimates.  However, the initial mass values
  are different, being larger for 47 Tuc. This result is an indication
  of the dependence of IFMR with the metallicity.

\end{itemize}

In future works  an extension to
higher initial masses should  be performed, from the theoretical point
of  view in  order  to  study younger  stellar  populations, as  open
clusters. Also,  this could be  used to constrain the  free parameters
governing the theoretical  mass loss rates that are  currently used in
stellar evolutionary computations.

\section*{Acknowledgments}

Part of this work was supported by  CNPq-Brazil and FAPERGS-Pronex Brazil. 
 We thank the anonymous referee for their useful comments and suggestions. 
A.D. Romero thanks L. Althaus, D. Koester and J. Isern for very 
useful discussion on this work. This research has made use  of NASA's 
Astrophysics Data System.     


\begin{thebibliography}{}

\bibitem{} Althaus, L.G., Serenelli, A.M., C{\'o}rsico, A.H., \& Montgomery, M.H.\ 2003, A\&A, 404, 593
\bibitem{} Althaus, L.~G., Serenelli, A.~M., Panei, J.~A., et al.\ 2005a, A\&A, 435, 631
\bibitem{} Althaus, L.~G., Miller Bertolami, M.~M., C{\'o}rsico, A.~H., Garc{\'{\i}}a-Berro, E., \& Gil-Pons, P.\ 2005b, A\&A, 440, L1 
\bibitem{} Althaus, L.~G., Garc{\'{\i}}a-Berro, E., Isern, J., C{\'o}rsico, A.~H., \& Rohrmann, R.~D.\ 2007, A\&A, 465, 249 
\bibitem{} Althaus, L.~G., C{\'o}rsico, A.~H., Bischoff-Kim, A., et al.\ 2010, ApJ, 717, 897
\bibitem{} Althaus, L.G., Garc{\'{\i}}a-Berro, E., Isern, J., C{\'o}rsico, A.H., \& Miller Bertolami, M.M.\ 2012, A\&A, 537, A33 
\bibitem{} Althaus, L.~G., Miller Bertolami, M.~M., \& C{\'o}rsico, A.~H.\ 2013, A\&A, 557, AA19 
\bibitem{} Angulo, C., et al. 1999, Nuclear Physics A, 656, 3%
\bibitem{} Basu, S., \& Antia, H.~M.\ 2008, Physics Reports, 457, 217 
\bibitem{} Bedin, L.~R., Salaris, M., Piotto, G., et al.\ 2009, ApJ, 697, 965 
\bibitem{} Bedin, L.~R., Salaris, M., King, I.~R., et al.\ 2010, ApJL, 708, L32 
\bibitem{} Bergeron, P., Saffer, R.~A., \& Liebert, J.\ 1992, ApJ, 394, 228 
\bibitem{} Bono, G., Salaris, M., \& Gilmozzi, R.\ 2013, A\&A, 549, AA102 
\bibitem{} Burgers, J.~M.\ 1969, Flow Equations for Composite Gases, New York: Academic Press, 1969  
\bibitem{} Carretta, E., Bragaglia, A., Gratton, R., D'Orazi, V., \& Lucatello, S.\ 2009, A\&A, 508, 695 
\bibitem{} Cassisi, S., Potekhin, A.Y., Pietrinferni, A., Catelan, M., \& Salaris, M.\ 2007, ApJ, 661, 1094 
\bibitem{} Castanheira, B. G., Kepler, S. O. 2009, MNRAS, 396, 1709%
\bibitem{} Catal{\'a}n, S., Isern, J., Garc{\'{\i}}a-Berro, E., et al.\ 2008a, A\&A, 477, 213 
\bibitem{} Catal{\'a}n, S., Isern, J., Garc{\'{\i}}a-Berro, E., \& Ribas, I.\ 2008b, MNRAS, 387, 1693 
\bibitem{} Caughlan, G. R., Fowler, W. A., Harris, M. J., \& Zimmermann, B. A. 1985, At. Data Nucl. Data Tables, 32, 197%
\bibitem{} Claver, C.~F., Liebert, J., Bergeron, P., \& Koester, D.\ 2001, ApJ, 563, 987
\bibitem{} Doherty, C.~L., Gil-Pons, P., Lau, H.~H.~B., et al.\ 2014, MNRAS, 441, 582  
\bibitem{} Dominguez, I., Chieffi, A., Limongi, M., \& Straniero, O.\ 1999, ApJ, 524, 226 
\bibitem{} Ferguson, J.~W., Alexander, D.~R., Allard, F., et al.\ 2005, ApJ, 623, 585 
\bibitem{} Freytag, B., Ludwig, H.-G., \& Steffen, M.\ 1996, A\&A, 313, 497 
\bibitem{} Garcia-Berro, E., Hernanz, M., Mochkovitch, R., \& Isern, J.\ 1988a, A\&A, 193, 141
\bibitem{} Garcia-Berro, E., Hernanz, M., Isern, J., \& Mochkovitch, R.\ 1988b, NATURE, 333, 642 
\bibitem{} Garc{\'{\i}}a-Berro, E., Torres, S., Althaus, L.~G., et al.\ 2010, NATURE, 465, 194  
\bibitem{} Garc{\'{\i}}a-Berro, E., Torres, S., Althaus, L.~G., \& Miller Bertolami, M.~M.\ 2014, A\&A, 571, AA56 
\bibitem{} Grevesse, N., \& Sauval, A.~J.\ 1998, Space Science Reviews, 85, 161 
\bibitem{} Groenewegen, M.~A.~T., \& de Jong, T.\ 1993, A\& A, 267, 410 
\bibitem{} Groenewegen, M.~A.~T., Sloan, G.~C., Soszy{\'n}ski, I., \& Petersen, E.~A.\ 2009, A\&A, 506, 1277 
\bibitem{} Haft, M., Raffelt, G., \& Weiss, A.\ 1994, ApJ, 425, 222 
\bibitem{} Hansen, B.~M.~S. et al. 2002, ApJ, 547, L155
\bibitem{} Hansen, B.~M.~S., Richer, H.~B., Fahlman, G.~G., et al.\ 2004, ApJS, 155, 551 
\bibitem{} Hansen, B.~M.~S., Anderson, J., Brewer, J., et al.\ 2007, ApJ, 671, 380
\bibitem{} Hansen, B.~M.~S., Kalirai, J.~S., Anderson, J., et al.\ 2013, NATURE, 500, 51 
\bibitem{} Hekker, S., Basu, S., Stello, D., et al.\ 2011, A\&A, 530, AA100 
\bibitem{} Herwig, F., Bl\"ocker, T., Sch\"onberner, D., \& El Eid, M. 1997, A\&A, 324, L81%
\bibitem{} Herwig , F. 2000, A\&A, 360, 952
\bibitem{} Herwig, F., Freytag, B., Fuchs, T., Hansen, J. P., Hueckstaedt, R. M., Porter, D. H., Timmes, F. X., \& Woodward, P. R. 2007,{\it Why Galaxies Care About AGB Stars: Their Importance as Actors and Probes}, 378, 43%
\bibitem{} Horowitz, C.J., Schneider, A.S., \& Berry, D.K.\ 2010, Physical Review Letters, 104, 231101 
\bibitem{} Hughto, J., Horowitz, C.J., Schneider, A.S., et al.\ 2012, Physical Review E, 86, 066413 
\bibitem{} Iben, I., Jr., \& MacDonald, J.\ 1986, ApJ, 301, 164 
\bibitem{} Iglesias, C.A., \& Rogers, F.J.\ 1996, ApJ, 464, 943 
\bibitem{} Isern, J., Garc{\'{\i}}a-Berro, E., Hernanz, M., Mochkovitch, R.,\& Torres, S.\ 1998, ApJ, 503, 239 
\bibitem{} Isern, J., Garc{\'{\i}}a-Berro, E., \& Salaris, M.\ 2001, Astrophysical Ages and Times Scales, 245, 328
\bibitem{} Isern, J., Garc{\'{\i}}a-Berro, E., Dom{\'{\i}}guez, I., Salaris, M., \& Straniero, O.\ 2005, 14th European Workshop on White Dwarfs, 334, 43 
\bibitem{} Itoh, N., Hayashi, H., Nishikawa, A., \& Kohyama, Y.\ 1996, ApJS, 102, 411 
\bibitem{} Kalirai, J.~S., Richer, H.~B., Fahlman, G.~G., et al.\ 2001, AJ, 122, 266 
\bibitem{} Kalirai, J.~S., Richer, H.~B., Reitzel, D., et al.\ 2005, ApJL, 618, L123 
\bibitem{} Kalirai, J.~S., Bergeron, P., Hansen, B.~M.~S., et al.\ 2007, ApJ, 671, 748 
\bibitem{} Kalirai, J.~S., Hansen, B.~M.~S., Kelson, D.~D., et al.\ 2008, ApJ, 676, 594 
\bibitem{} Kalirai, J.~S., Saul Davis, D., Richer, H.~B., et al.\ 2009, ApJ, 705, 408 
\bibitem{} Kalirai, J.~S.\ 2013, MmSAI, 84, 58
\bibitem{} Kalirai, J.~S., Marigo, P., \& Tremblay, P.-E.\ 2014, ApJ, 782, 17 
\bibitem{} Karakas, A., \& Lattanzio, J.~C.\ 2007, PASA, 24, 103 
\bibitem{} Karakas, A.~I.\ 2010, MNRAS, 403, 1413 
\bibitem{} Kepler, S.~O., Pelisoli, I., Koester, D., et al.\ 2014, arXiv:1411.4149  
\bibitem{} Kleinman, S.~J., Kepler, S.~O., Koester, D., et al.\ 2013, ApJS, 204, 5 
\bibitem{} Krauss, L.~M., \& Chaboyer, B.\ 2003, Science, 299, 65 
\bibitem{} Kurtz, D.~W., Shibahashi, H., Dhillon, V.~S., et al.\ 2013, MNRAS, 432, 1632 
\bibitem{} Liebert, J., Bergeron, P., \& Holberg, J.~B.\ 2005, ApJS, 156, 47 
\bibitem{} Lambert, D.~L., Gustafsson, B., Eriksson, K., \& Hinkle, K.~H.\ 1986, ApJS, 62, 373 
\bibitem{} Lugaro, M., Herwig, F., Lattanzio, J. C., Gallino, R., \& Straniero, O. 2003, ApJ, 586, 1305%
\bibitem{} Magni, G., \& Mazzitelli, I.\ 1979, A\&A, 72, 134 
\bibitem{} Marigo, P. \ 2002, A\& A, 387, 507 
\bibitem{} Marigo, P., \& Girardi, L.\ 2007, A\&A, 469, 239 
\bibitem{} Marigo, P., \& Aringer, B.\ 2009, A\&A, 508, 1539 
\bibitem{} Marigo, P., Bressan, A., Nanni, A., Girardi, L., \& Pumo, M.~L.\ 2013, MNRAS, 434, 488
\bibitem{} Mazzitelli, I., D'Antona, F., \& Ventura, P.\ 1999, A\&A, 348, 846 
\bibitem{} Medin, Z., \& Cumming, A.\ 2010, Physical Review E, 81, 036107 
\bibitem{} Miglio, A., Brogaard, K., Stello, D., et al.\ 2012, MNRAS, 419, 2077 
\bibitem{} Miller Bertolami, M.~M., Althaus, L.~G., Unglaub, K., \& Weiss, A.\ 2008, A\&A, 491, 253 
\bibitem{} Miller Bertolami, M.~M., Althaus, L.~G., \& Garc{\'{\i}}a-Berro, E.\ 2013, ApJL, 775, L22 
\bibitem{} Moehler, S., Koester, D., Zoccali, M., et al.\ 2004, A\&A, 420, 515 
\bibitem{} Monelli, M., Corsi, C.~E., Castellani, V., et al.\ 2005, ApJL, 621, L117 
\bibitem{} Mucciarelli, A., Salaris, M., Lovisi, L., et al.\ 2011, MNRAS, 412, 81 
\bibitem{} Ohnaka, K., Tsuji, T., \& Aoki, W.\ 2000, A\&A, 353, 528 
\bibitem{} Paxton, B., Bildsten, L., Dotter, A., et al.\ 2011, ApJS, 192, 3 
\bibitem{} Paxton, B., Cantiello, M., Arras, P., et al.\ 2013, ApJ, 208, 4 
\bibitem{} Perryman, M.~A.~C., Brown, A.~G.~A., Lebreton, Y., et al.\ 1998, A\&A, 331, 81 
\bibitem{} Prada Moroni, P.~G., \& Straniero, O.\ 2002, ApJ, 581, 585 
\bibitem{} Reimers, D.\ 1975, MSRSL, 8, 369 
\bibitem{} Renedo, I., Althaus,  L. G., Miller Bertolami, M. M., Romero, A. D., C\'orsico, A. H.,  Rohrmann, R. D., \& Garc\'ia-Berro, E. 2010, ApJ, 717, 183%
\bibitem{} Renzini, A., \& Fusi Pecci, F.\ 1988, ARA\&A, 26, 199
\bibitem{} Renzini, A., Bragaglia, A., Ferraro, F.~R., et al.\ 1996, ApJL, 465, L23 
\bibitem{} Richer, H.~B., Fahlman, G.~G., Brewer, J., et al.\ 2004, AJ, 127, 2771 
\bibitem{} Richer, H.~B., Hansen, B.~M., Davis, S., et al.\ 2006, Bulletin of the American Astronomical Society, 38, \#228.01
\bibitem{} Rohrmann, R.~D., Althaus, L.~G., Garc{\'{\i}}a-Berro, E., C{\'o}rsico, A.~H., \& Miller Bertolami, M.~M.\ 2012, A\&A, 546, A119 
\bibitem{} Romero, A.~D., C{\'o}rsico, A.~H., Althaus, L.~G., et al.\ 2012, MNRAS, 420, 1462 
\bibitem{} Romero, A.~D., Kepler, S.~O., C{\'o}rsico, A.~H., Althaus, L.~G., \& Fraga, L.\ 2013, ApJ, 779, 58 
\bibitem{} Salaris, M., Dom\'inguez, I., Garc\'ia-Berro, E., Hernanz, M., Isern, J., Mochkovitch, R.  1997,  ApJ, 486, 413%
\bibitem{} Salaris, M., Serenelli, A., Weiss, A., \& Miller Bertolami, M. 2009, ApJ, 692, 1013 
\bibitem{} Salaris, M., Althaus, L.~G., \& Garc{\'{\i}}a-Berro, E.\ 2013, A\&A, 555, AA96 
\bibitem{} Schaller, G., Schaerer, D., Meynet, G., \& Maeder, A.\ 1992, A\&AS, 96, 269
\bibitem{} Schneider, A.S., Hughto, J., Horowitz, C.J., \& Berry, D.K.\ 2012, Physical Review E, 85, 066405 
\bibitem{} Schr\"oder, K. P., \& Cuntz, M. 2005, ApJL, 630, L73
\bibitem{} Segretain, L., \& Chabrier, G.\ 1993, A\&A, 271, L13
\bibitem{} Segretain, L., Chabrier, G., Hernanz, M., et al.\ 1994, ApJ, 434, 641%
\bibitem{} Smartt, S. J. 2009, ARA\&A, 47, 63
\bibitem{} Stancliffe, R.~J., Izzard, R.~G., \& Tout, C.~A.\ 2005, MNRAS, 356, L1 
\bibitem{} Straniero, O., Dom\'{\i}nguez, I., Imbriani, G., \& Piersanti, L. 2003, ApJ, 583, 878%
\bibitem{} Tassoul, M., Fontaine, G., Winget, D. E. 1990, ApJS, 72, 335%
\bibitem{} Torres, S., Garc{\'{\i}}a-Berro, E., Burkert, A., \& Isern, J.\ 2002, MNRAS, 336, 971 
\bibitem{} Tremblay, P.-E., \& Bergeron, P. 2008, ApJ, 672, 1144%
\bibitem{} van Horn, H.M.\ 1968, ApJ, 151, 227 
\bibitem{} Vassiliadis, E.  \& Wood, P. R. 1993, ApJ, 413, 641%
\bibitem{} Wachlin, F.~C., Miller Bertolami, M.~M., \& Althaus, L.~G.\ 2011, A\&A, 533, AA139 
\bibitem{} Weiss, A., \& Ferguson, J.~W.\ 2009, A\&A, 508, 1343 
\bibitem{} Weidemann, V.\ 1977, A\& A, 59, 411
\bibitem{} Weidemann, V.\ 2000, A\&A, 363, 647 
\bibitem{} Winget, D.~E., Hansen, C.~J., Liebert, J., et al.\ 1987, ApJL, 315, L77 
\bibitem{} Winget, D.E., Kepler, S.O., Campos, F., et al.\ 2009, ApJL, 693, L6 
\bibitem{} Winget, D.~E., Montgomery, M.~H., Kepler, S.~O., Campos, F., \& Bergeron, P.\ 2010, AIPCS, 1273, 146 
\bibitem{} Wu, Z.-Y., Zhou, X., Ma, J., et al.\ 2007, AJ, 133, 2061 
\end{thebibliography}
\end{document}